\documentclass[]{aa}
\usepackage[colorlinks=true,linkcolor=blue,citecolor=blue]{hyperref}
\usepackage{lipsum,multicol}
\usepackage{graphicx}

\usepackage{natbib}

\def \kms{km\,s$^{-1}$}
\def \ms{m\,s$^{-1}$}

\def \mjup{$M_\mathrm{Jup}$}

\def \sun{\odot}
\def \Ha{H$\alpha$\,\,}

\bibpunct{(}{)}{;}{a}{}{,} % to follow the A&A style
\usepackage{txfonts}
\usepackage{soul}
\usepackage{pdflscape}
\usepackage{capt-of}
\usepackage{placeins}
\usepackage{hyperref}
\usepackage{tablefootnote}
\usepackage{tikz}
\DeclareMathAlphabet{\mathpzc}{OT1}{pzc}{m}{it}
\def\CrI{\hbox{\rm Cr\,$\scriptstyle\rm I$}}
\def\FeI{\hbox{\rm Fe\,$\scriptstyle\rm I$}}
\def\NaI{\hbox{\rm Na\,$\scriptstyle\rm I$}}
\def\MgI{\hbox{\rm Mg\,$\scriptstyle\rm I$}}
\def\LiI{\hbox{\rm Li\,$\scriptstyle\rm I$}}

\def\CI{\hbox{\rm C\,$\scriptstyle\rm I$}}
\def\BaI{\hbox{\rm Ba\,$\scriptstyle\rm I$}}
\def\CaI{\hbox{\rm Ca\,$\scriptstyle\rm I$}}

\def\HI{\hbox{\rm H\,$\scriptstyle\rm I$}}

\def\NiI{\hbox{\rm Ni\,$\scriptstyle\rm I$}}
\def\WI{\hbox{\rm W\,$\scriptstyle\rm I$}}

\def\CaII{\hbox{\rm Ca\,$\scriptstyle\rm II$}}
\def\BaII{\hbox{\rm Ba\,$\scriptstyle\rm II$}}
\def\FeII{\hbox{\rm Fe\,$\scriptstyle\rm II$}}
\def\VII{\hbox{\rm V\,$\scriptstyle\rm II$}}
\def\CrII{\hbox{\rm Cr\,$\scriptstyle\rm II$}}
\def\SrII{\hbox{\rm Sr\,$\scriptstyle\rm II$}}
\def\YII{\hbox{\rm Y\,$\scriptstyle\rm II$}}

\def\TiII{\hbox{\rm Ti\,$\scriptstyle\rm II$}}
\def\NdII{\hbox{\rm Nd\,$\scriptstyle\rm II$}}
\def\EuII{\hbox{\rm Eu\,$\scriptstyle\rm II$}}
\def\ScII{\hbox{\rm Sc\,$\scriptstyle\rm II$}}

\begin{document}

   \title{The GAPS programme at TNG}

   \subtitle{LXXXIII. Orbital refinement and atmospheric investigation \\ of TOI-1518 b (BD+66 1610 b)}

   \author{M.\,C.~D'Arpa\inst{1} \thanks{Corresponding author: mattia.darpa@inaf.it}
          \and
          F.~Biassoni\inst{2, 3}
          \and 
          F.~Borsa \inst{2}
          \and
          V.~Singh \inst{4}
          \and
          G.~Micela \inst{1}
          \and 
          S.~Benatti \inst{1}
          \and
          L.~Malavolta \inst{5, 6}
          \and
          A.\,F.~Lanza \inst{4}
          \and
          S.~Desidera \inst{6}
          \and
          E.~Herrero-Cisneros \inst{7}
          \and
          A.\,S.~Bonomo \inst{8}
          \and
          M.~Brogi \inst{8, 9}
          \and
          G.~Guilluy \inst{8}
          \and
          L.~Fossati \inst{10}
          \and 
          L.~Mancini \inst{8, 11}
          \and 
          G.~Mantovan \inst{5, 12}
          \and
          J.~Sanz-Forcada \inst{7}
          \and
          A.~Sozzetti \inst{8}
          \and
          M.\,R.~Zapatero Osorio \inst{7}
          \and
          V.~Fardella \inst{1,13}
          \and
          A.~Ghedina \inst{14}
          \and
          A.~Harutyunyan \inst{14}
          \and
          G.~Scandariato \inst{4}
          \and
          T.~Zingales \inst{5, 6}
          }

   \institute{INAF -- Palermo Astronomical Observatory, Piazza del Parlamento, 1, 90134 Palermo, Italy %1
         \and
            INAF -- Osservatorio Astronomico di Brera, via E. Bianchi 46, 23807 Merate, Italy %2
        \and
            Dipartimento di Scienza e Alta Tecnologia, Universitá degli Studi dell’Insubria, via Valleggio 11, 22100 Como, Italy %3
        \and
            INAF -- Osservatorio Astrofisico di Catania, Via S. Sofia 78, 95123 Catania, Italy %4
        \and
            Dipartimento di Fisica e Astronomia “Galileo Galilei”, Università di Padova, Vicolo dell’Osservatorio 3, 35122 Padova, Italy %5
        \and
            INAF -- Osservatorio Astronomico di Padova, Vicolo dell’Osservatorio 5, 35122 Padova, Italy %6
        \and
            Centro de Astrobiología, CSIC-INTA, Camino Bajo del Castillo s/n, 28692 Villanueva de la Cañada, Madrid, Spain %7
        \and
            INAF -- Osservatorio Astrofisico di Torino, Via Osservatorio 20, 10025 Pino Torinese, Italy %8
        \and
            Dipartimento di Fisica, Università di Torino, Via Pietro Giuria 1, 10125 Torino, Italy %9
        \and
            Institut für Weltraumforschung, Austrian Academy of Sciences, Schmiedlstrasse 6, 8042 Graz, Austria %10
        \and
            Dipartimento di Fisica, Università di Roma “Tor Vergata”, Via della Ricerca Scientifica 1, 00133 Roma, Italy %11
        \and
            Centro di Ateneo di Studi e Attività Spaziali “G. Colombo”, Università di Padova, Via Venezia 15, 35131 Padova %12
        \and
            Università di Palermo, Dipartimento di Fisica e Chimica Emilio Segrè, Via Archirafi 36, Palermo, Italy %13
        \and
            Fundación Galileo Galilei--INAF, Rambla José Ana Fernandez Pérez 7, 38712 Breña Baja, Tenerife, Spain %14
        }
   \date{Received 29 July 2026 / Accepted dd November 2026}

% \abstract{}{}{}{}{} 
% 5 {} token are mandatory
 
 \abstract
   {TOI-1518 b (BD+66 1610 b) is a highly irradiated ultra-hot Jupiter on a strongly misaligned orbit, making it a benchmark target for atmospheric and dynamical studies.}
   {We aim to comprehensively characterise the TOI-1518 system by refining its orbital and physical parameters, and by investigating the chemical composition of the planetary atmosphere through transmission spectroscopy.}
   {We employed a joint Bayesian framework combining TESS photometry, HARPS-N high-resolution spectroscopy, and least-squares deconvolution profiles to break stellar geometrical degeneracies. We then applied single-line and cross-correlation function (CCF) transmission spectroscopy across theoretical templates.}
   {By anchoring the stellar projected rotational velocity ($v \sin i_\star \approx 79.7$\,\kms), we confirmed the nearly polar orbit ($\lambda \approx 238$ deg) and the theoretically predicted nodal precession ($\mathrm{d}i/\mathrm{d}t \approx 0.15$ deg yr$^{-1}$). We revealed an atmospheric Rossiter-McLaughlin (RM) effect and a bulk atmospheric blue-shift of $\approx -7.0$\,\kms. Through transmission spectroscopy, we detected \HI, \CI, \NaI, \CaI, \ScII, \TiII, \VII, \CrI, \FeI, \FeII, \NiI, \SrII, \YII, \EuII, and \WI. This marks the first detection of \HI, \CI, \ScII, \VII, \SrII, \YII, \EuII, and \WI \ in this planet, and the first ever detection of \EuII, and \WI \ in an exoplanetary atmosphere. Inconsistencies in the \NaI\ doublet shifts suggest likely stellar contamination. Furthermore, time-resolved CCF analysis revealed morning-to-evening limb asymmetries for species like \MgI\, and \BaI, hinting at night-side cold traps.}
   {The atmosphere of TOI-1518 b exhibits a rich chemical inventory shaped by intense irradiation and bulk circulation. Our results emphasize the necessity of simultaneously modelling stellar line morphology and planetary tracks to accurately interpret atmospheric signals under strong RM effects.}

    
   \titlerunning{Orbital refinement and atmospheric investigation of TOI-1518 b (BD+66 1610 b)}
   \authorrunning{M. C. D'Arpa et al.}
    
   \maketitle
%
%-------------------------------------------------------------------

\section{Introduction}
One of the primary objectives of exoplanetary research for the upcoming decade is the atmospheric characterisation of terrestrial planets, with the ultimate goal of identifying potential biosignatures \citep{Fujii2018}. High-resolution transmission spectroscopy (HRTS) has emerged as one of the most promising techniques with the potential to achieve this milestone. By resolving individual spectral lines, HRTS allows for the unambiguous identification of molecular and atomic species. It leverages the distinct Doppler shift of the moving planet to effectively disentangle the planetary signal from the dominant, relatively static stellar and telluric spectra, typically employing the cross-correlation function (CCF) technique \citep[e.g.,][]{Snellen_2010, Brogi_2012}. 
While the application of these advanced techniques to progressively cooler and smaller planets is of immense scientific interest and offers groundbreaking potential, pushing these limits requires extreme precision and a robust understanding of instrumental and astrophysical systematics. Therefore, to rigorously verify and corroborate current analysis frameworks and to develop novel methodologies, targets that provide high signal to noise ratio (S/N) observable features are essential. In this context, ultra-hot Jupiters (UHJs) serve not only as ideal astrophysical benchmarks \citep[e.g.,][]{hoeijmakers2018atomic}, but also as profoundly compelling targets in their own right. Their extreme temperatures significantly expand the atmospheric pressure scale height, boosting the amplitude and detectability of spectral features. Furthermore, by preventing the condensation of obscuring clouds and hazes, UHJs offer an unhindered view of their atmospheric bulk composition. Characterizing this pristine chemical inventory is intrinsically crucial, as it provides fundamental clues regarding the formation mechanisms and early dynamical evolution of gas giant planets.
Defined by equilibrium temperatures exceeding 2000~K \citep{Fortney2008}, UHJs are characterised by extreme environments in which most molecules are thermally dissociated. Consequently, their upper atmosphere hosts vaporised metals, both neutral and ionised \citep[e.g.,][]{Casasayas-Barris2017, Hoeijmakers2019}. These metals, along with the high-temperature molecules containing them, are recognised as strong sources of opacity in the optical and near-ultraviolet regions \citep{Fortney2008, Lothringer2018}. The absorption of intense stellar irradiation by these species is predicted to be the primary driver for the formation of strong thermal inversions (stratospheres) often exhibited by UHJs \citep[e.g.,][]{hubeny2003, haynes_2015}. Nevertheless, the exact chemical species responsible for these inversions remain a subject of active debate \citep{Fortney2008, Lothringer2022}, necessitating precise observational constraints. Furthermore, in these extreme regimes, intense UV irradiation can drive the upper atmosphere out of local thermodynamic equilibrium (NLTE), significantly altering the level populations of atomic species and complicating the interpretation of line strengths \citep{Fossati2021}.

Beyond the detection of important species, HRTS serves as a vital means of probing the complex dynamics of UHJ atmospheres. The analysis of high-resolution line profiles allows for the characterisation of atmospheric kinematics, including day-to-night side winds and equatorial jets \citep{darpa_kelt9, borsa_rot,  2019A&A...628A...9C}, and provides insights into extended atmospheres and potential escape mechanisms \citep{borsa_ox_nature}. However, applying HRTS to UHJs often involves navigating significant astrophysical challenges. The planetary transit across a rotating stellar disk creates a well-known spectral distortion referred to as the Doppler shadow, or the Rossiter-McLaughlin (RM) effect \citep{CollierCameron2010}. While this phenomenon occurs in all rotating systems, it becomes particularly challenging for UHJs, which frequently orbit early-type, fast-rotating hosts. In these extreme regimes, the highly broadened Doppler shadow often strongly overlaps with the planetary track in velocity space, making the disentanglement of atmospheric absorption from stellar surface variations a critical task. However, this signal is not solely a source of contamination. In this work, rather than treating the RM effect merely as a nuisance, we exploit it as a valuable resource to independently refine the orbital parameters of the system.

Among the most compelling targets for pushing the capabilities of HRTS is TOI-1518 b (BD+66 1610 b), a highly irradiated gas-giant planet possessing iron vapour in its atmosphere \citep{cabot21}. Discovered by the Transiting Exoplanet Survey Satellite \citep[TESS,][]{ricker2015transiting}, this planet orbits a fast-rotating ($v \sin i_\star = 85.1 \pm 6.0$~\kms) F0\,V-type star with an effective temperature of approximately $7300$~K and a $V$ magnitude of $\sim$ 8.95. It is among the first exoplanets discovered by TESS to yield a high-resolution detection of an atmospheric species, as demonstrated by \citet{cabot21}. Table \ref{tab:toi1518_par} summarises the physical properties of TOI-1518 and TOI-1518 b. 

With an equilibrium temperature of $T_{\text{eq}} \approx 2580$~K, TOI-1518 b resides in a thermal regime where most molecules are thermally dissociated, leading to a rich expected inventory of vaporised atomic and ionic species. While its F0-type host star implies a significantly softer UV irradiation environment compared to the A0-type host of the archetypal UHJ KELT-9 b, which inherently drives different photochemical and ionization balances, the comparable temperature makes it an ideal candidate for a strictly analogous application of single-line and cross-correlation analyses. Furthermore, the system parameters result in a remarkably large atmospheric scale height ($H \approx 609$~km, corresponding to $\sim 0.48\%$ of the planetary radius) and a transmission spectroscopy metric (TSM) of 352. Since the best gas giants for transmission spectroscopy typically have a TSM larger than 90 \citep{kempton2018}, this underlines the exceptional potentiality of TOI-1518 b for detailed atmospheric characterisation.

Since its discovery, TOI-1518 b has garnered significant attention, rapidly becoming a benchmark target for both dynamical and atmospheric studies. The initial characterisation by \citet{cabot21} not only confirmed the presence of neutral and ionised iron in its upper atmosphere but also revealed a highly misaligned, near-polar orbit with a grazing transit impact parameter.  This peculiar orbital architecture (shown in Fig. \ref{fig:transit_geometry}), coupled with the oblate shape of the fast-rotating host star, drives a measurable nodal precession of the planetary orbit, which was recently detected by \citet{Watanabe2024}. 

Beyond its orbital dynamics, the atmosphere of TOI-1518 b has been the subject of intensive high-resolution spectroscopic campaigns. Recent observations with the MAROON-X spectrograph by \citet{Simonnin2025} resolved the time-dependent absorption of multiple chemical species, revealing evidence of a strong atmospheric drag that significantly slows the planetary winds, with distinct kinematic signatures between neutral and ionised species. In parallel, the PEPSI Investigation, Retrieval, and Atlas of Numerous Giant Atmospheres (PIRANGA) survey utilised the Potsdam Echelle Polarimetric and Spectroscopic Instrument (PEPSI) at the Large Binocular Telescope to further constrain the atmospheric composition and wind profiles \citep{Basinger2025}, while high-resolution emission spectroscopy has independently confirmed the presence of a strong thermal inversion on the planet's dayside \citep{Petz2025}. Building on these kinematic discrepancies, \cite{2026NatAs..10.1145S} recently demonstrated that this drag is magnetic in nature, exploiting the velocity offset between ions and neutrals to provide an estimate of TOI-1518b's planetary magnetic field.

Building upon this rich and rapidly evolving literature, our work provides a significant step forward by introducing a fully self-consistent framework that directly bridges the system's orbital dynamics with its atmospheric physics. Rather than treating these aspects separately, we implement a global joint Bayesian analysis combining TESS photometry, HARPS-N radial velocities (RVs), and tomographic least-squares deconvolution (LSD) profiles. This approach allows us to break geometric degeneracies, refining the planetary mass and system architecture while isolating the kinematic signature of the atmospheric RM effect. Furthermore, we leverage our extensive observational baseline to perform a comprehensive individual-line and cross-correlation atmospheric survey, enabling not only the identification of a rich chemical inventory but also a time-resolved investigation of chemical asymmetries between the morning and evening limbs of the planet.

The paper is organised as follows.
In Sect. \ref{sec:observations}, we describe the datasets we analysed. In Sect. \ref{sec:analysis_results}, we detail the joint analytical framework used to refine the system parameters and present our findings regarding the atmospheric characterisation of the target. Finally, we discuss our results in the context of the current literature in Sect. \ref{sec:discussion}, and outline our main conclusions in Sect. \ref{sec:conclusions}.
%--------------------------------------------------------------------

\section{Observations}
\label{sec:observations}
We analysed four transits of the planet collected from 26 August 2023 to 7 October 2023 with the high-resolution (R $\sim$ 115\,000) HARPS-N spectrograph (3800--6900 $\AA$) \citep{2012SPIE.8446E..1VC} located at Telescopio Nazionale Galileo (TNG). The first two nights were collected by CAT23A\_69 programme (P.I. Herrero Cisneros), while the last two were obtained as part of the BRIDGES large programme (P.I. Borsa) within the Italian GAPS collaboration. 
A complete overview of the observations, including the date, programme number, PI, number of in- and out-of-transit spectra, average S/N and airmass range is given in Table \ref{tab:observations}. We also show the S/N and the airmass as a function of the orbital phase in Fig. \ref{fig:toi1518_obs}. The night reports for the different nights do not give any important observing conditions to remark. For all the analyses discussed in this work we decided to discard all the spectra with a S/N smaller than 40 resulting in 21 discarded spectra out of 160. 

\section{Analysis and results}
\label{sec:analysis_results}
In the following section, we describe the two main components of our study: the refinement of the system's orbital architecture and the characterisation of the planetary atmosphere. We first present the joint modelling of the photometric and spectroscopic datasets to update the system parameters. Subsequently, we detail the analysis of the atmospheric signatures through single line analysis and cross correlation with templates.
\subsection{Refinement of the orbital parameters}
\label{sec:parameters}
An accurate knowledge of the system's orbital and physical parameters is a fundamental prerequisite for HRTS. The planetary RV semi-amplitude ($K_{\text{p}}$) is strictly required to correctly shift the observed spectra and map the atmospheric cross-correlation signal into the planetary rest frame. Furthermore, for targets orbiting fast-rotating stars like TOI-1518, an exact characterisation of the stellar projected rotational velocity ($v \sin{i_{\star}}$) and the projected spin-orbit angle ($\lambda$) is critical to accurately model and isolate the Doppler shadow from the underlying planetary atmospheric absorption. 

Previous studies by \citet{cabot21} and \citet{Watanabe2024} established a robust foundational understanding of the TOI-1518 system, revealing a highly misaligned, near-polar orbit and providing initial constraints on the stellar and planetary parameters. However, to maximise the S/N of our atmospheric detections and to fully exploit the extensive baseline of our observational campaign, we sought to further refine these orbital elements. Crucially, \citet{Watanabe2024} theoretically predicted a rapid nodal precession for this system; thus, a robust refinement of the transit geometry strictly requires accounting for a time-dependent orbital inclination.

To achieve the highest possible precision and break known parameter degeneracies, we performed a comprehensive joint global fit. Our modelling architecture simultaneously incorporates photometric and high-resolution spectroscopic datasets. Specifically, we performed a joint analysis combining the available TESS transit light curves, RVs extracted from our HARPS-N spectra, and the spectroscopic variations mapped through the residuals of the LSD profiles. Furthermore, we directly incorporated the modelling of the master out-of-transit LSD profile into our Markov chain Monte Carlo (MCMC) framework, fitting it with a \citet{Gray2005} rotational broadening function. This inclusion is critical to simultaneously and self-consistently constrain the stellar line profile parameters, most notably the $v \sin i_\star$, alongside the planetary parameters. Ultimately, this synergistic approach allows us to tightly couple the geometric constraints provided by the photometric transit with the dynamical information encoded in both the time-resolved RM effect and the intrinsic stellar line morphology. The step-by-step procedures regarding the extraction of these datasets and the exact setup of the Bayesian joint modelling framework are detailed in the following subsections.

\subsubsection{TESS light curve preparation}
\label{sec:tess}
TESS \citep{ricker2015transiting} observed TOI-1518 across five sectors: Sectors 57 and 58 (from 30 September to 26 November 2022), Sector 77 (from 26 March to 23 April 2024), and Sectors 84 and 85 (from 1 October to 21 November 2024). We extracted these data using the {\tt\string Lightkurve} Python package \citep{LightKurve2018ascl.soft12013L}, which facilitates the retrieval of data products from the Mikulski Archive for Space Telescopes (MAST). For our analysis, we specifically utilised the 120-s cadence pre-search data conditioning simple aperture photometry PDC-SAP light curves. These products are processed via the Science Processing Operations Center SPOC pipeline \citep{Jenkins2016, Jenkins2019ESS.....430202J, Jenkins2020DPS....5220703J, Jenkins2021tsc2.confE.183J}, which mitigates instrumental systematics while preserving astrophysical variability.
While the discovery paper by \cite{cabot21} incorporated 1800-second long-cadence data from Sectors 17 and 18, we chose to exclude such earlier observations from our analysis. We determined that the superior temporal resolution and high S/N provided by the five subsequent sectors of 120-s cadence data were enough to achieve the objectives of this work. 
Prior to our analysis, we filtered the dataset by removing all observations flagged with bad quality bitmasks and NaN values, followed by a $3.5\sigma$ outlier clipping.
This refinement resulted in the removal of only a few tens of data points in total, ensuring a clean and reliable light curve for our subsequent analysis.

\subsubsection{HARPS-N RVs and LSD profiles extraction}

To extract high-fidelity average line profiles from our high-resolution spectra, we employed the LSD technique \citep{Donati1997}. While the CCF is routinely employed in exoplanetary literature to compute average line profiles and derive precise RVs, it mathematically produces a cross-correlation peak rather than a true physical absorption line. Consequently, the CCF profile is inherently broadened by the cross-correlation process itself and suffers from significant line-blending artifacts when applied to fast-rotating stars. In contrast, LSD solves a linear inverse problem to deconvolve the observed spectrum from a theoretical line list. This approach effectively disentangles blended lines and perfectly preserves the true physical shape and macroscopic broadening of the average photospheric line. Given the rapid rotation of the F0-type host star TOI-1518, preserving the exact morphology of the line profile is mandatory to accurately model the rotational broadening and the RM effect.

The normalisation on 1D stellar spectra (see Sect. \ref{sec:atmo} for their initial pre-processing and extraction) was performed using the {\tt\string PARVATI}\footnote{\url{https://pypi.org/project/parvati/}} pipeline (Rainer in prep.). For the LSD computation, we generated a custom stellar line mask utilizing the VALD3 database \citep{Ryabchikova2015}. The mask was specifically tailored to the fundamental parameters of TOI-1518 ($T_{\text{eff}} \approx 7300$~K, $\log{g}$ $\approx 4.1$, and $[\text{Fe/H}] \approx -0.1$), selecting only absorption lines with suitable depths to optimize the S/N of the resulting profiles.

Following the extraction of the LSD profiles, we measured the stellar RVs for each out-of-transit exposure to trace the Keplerian reflex motion. Because standard Gaussian fitting is inadequate for such a rapidly rotating star, we modelled the profiles using the rotational broadening function described by \citet{Gray2005}. 

Subsequently, to prepare the data for the joint orbital modelling and the tomographic analysis, we computed a high-S/N master out-of-transit LSD profile. Rather than extracting a static value of the $v \sin i_\star$ at this stage, we integrated this master profile directly into our MCMC joint fit. By simultaneously fitting the master LSD profile alongside the photometric and RV data using the \citet{Gray2005} function, we allow the MCMC to self-consistently constrain $v \sin i_\star$, the limb-darkening coefficients, and the systemic velocity, fully propagating their uncertainties into the derived planetary parameters.

Finally, to isolate the planetary transit signature, the Doppler shadow, from the static stellar absorption, we divided each individual in-transit LSD profile by the corresponding master out-of-transit profile. The resulting residual profiles were then shifted into the stellar rest frame and sorted by their orbital phase. This procedure effectively removes the static stellar lines and quasi-static instrumental effects, yielding a 2D phase-resolved tomographic map. In this 2D map shown in Fig. \ref{fig:mcmc_map}, the time-dependent spectral distortion caused by the transiting planet blocking the rotating stellar hemisphere emerges clearly, providing the dynamic RM input for our joint orbital refinement.

\subsubsection{Joint Bayesian fit setup}
\label{sec:joint_fit}

To derive the posterior probability distributions of the system parameters and properly account for their covariances, we performed a global Bayesian analysis utilizing the affine-invariant MCMC ensemble sampler implemented in the {\tt\string emcee} Python package \citep{ForemanMackey2013}. 

Our global log-likelihood is constructed as the sum of the independent log-likelihoods from four distinct observational datasets: $\ln \mathcal{L}_{\text{global}} = \ln \mathcal{L}_{\text{TESS}} + \ln \mathcal{L}_{\text{RV}} + \ln \mathcal{L}_{\text{MAP}} + \ln \mathcal{L}_{\text{PROF}}$. We assumed Gaussian errors for all datasets. We initialised 48 walkers, drawing their starting positions from uniform distributions bounded by strictly physical limits. We applied Gaussian or uniform priors to the free parameters, centred on the values derived from our preliminary independent analyses and previous literature. To rigorously ensure the convergence of our sampling, we monitored the integrated autocorrelation time ($\tau$) of the chains dynamically during the run. The MCMC exploration was deemed converged, and the burn-in phase concluded, only when the total chain length exceeded $50\tau$ and the fractional change of $\tau$ between consecutive evaluations dropped below $1\%$. 

To ensure the stability of the MCMC sampling, we performed a preliminary fit following the photometric modelling methodology detailed in previous studies \citep[e.g.,][]{Scandariato2022, Singh2024}, which couples analytical transit models with baseline de-trending to derive initial guesses and priors of the orbital period ($P$) and the mid-transit time ($T_c$). Finally, we calculated the linear limb-darkening coefficients for both datasets using the \texttt{ExoCTK} toolkit\footnote{\url{https://exoctk.stsci.edu/}}, employing these values as priors in our analysis. For the TESS photometry, we adopted the coefficients corresponding to the standard TESS bandpass, $T$. For the HARPS-N spectroscopic data, we computed the coefficients by employing a custom top-hat filter centred on the instrument's effective wavelength range, ensuring that the modelled line profiles were consistent with the spectroscopic LSD extraction.
The specific free parameters actively driving the physical model of each dataset within the joint framework are detailed below, while a summary of the physical parameters constrained by each observational dataset can be found in Table \ref{tab:parameter_matrix}.
\begin{itemize}
    \item TESS light curve (TESS): The transit photometry was modelled using the \texttt{batman} package \citep{Kreidberg2015}. To empirically test the nodal precession predicted by \citet{Watanabe2024}, we upgraded the standard transit model to allow the orbital inclination to vary linearly with time across the observed epochs. Consequently, this dataset strongly constrains the reference inclination at $T_c$ ($i_0$), its temporal derivative ($\mathrm{d}i/\mathrm{d}t$), and the core planetary-to-stellar radius ratio ($R_p/R_\star$).
    
    \item High resolution RVs (RV): The 1D RVs were modelled by combining a standard Keplerian circular orbit with the classical analytical RM model provided by the \texttt{PyAstronomy} suite \citep{pya}, augmented by the atmospheric RM effect framework as described by \citet{borsa_ar, borsa_rot}. The inclusion of this atmospheric contribution accounts for the additional stellar velocity field dynamically occulted by the extended planetary atmosphere. To optimize the efficiency and stability of the MCMC sampling, the absolute systemic velocity was not fitted directly. Instead, the RVs were pre-centred by subtracting the median of the out-of-transit points, allowing us to fit for a small residual velocity offset ($\gamma$). Throughout this work, we maintain this distinction: $V_\mathrm{sys}$ denotes the true physical systemic velocity of the star, while $\gamma$ refers exclusively to the fitted dataset-specific residual. Consequently, this dataset constrains the stellar RV semi-amplitude ($K$), $\gamma$, $v \sin i_\star$, and $\lambda$, while also directly measuring the effective atmospheric radius extension $(R_p/R_\star)_{\text{atmo}}$.
    
    \item 2D Doppler tomography (MAP): The time-resolved spectroscopic residuals were modelled by mapping the dynamic path of the occulted stellar surface (Doppler shadow) and the planetary atmospheric absorption. The geometric track is governed by $v \sin i_\star$, $\lambda$, and $i$. The planetary absorption track is strictly constrained by the $K_p$ and the net atmospheric wind velocity shift ($v_w$). Additionally, phenomenological Gaussian parameters ($a_1, \sigma_1$ for the shadow; $a_2, \sigma_2$ for the planet) were simultaneously fitted to model the depths and widths of the tomographic traces. In Appendix \ref{app:sigma}, we discuss the constraints imposed on the $\sigma_2$ parameter.
    
    \item Master out-of-transit profile (PROF): The high-S/N mean LSD profile was modelled using the analytical rotational broadening kernel described by \citet{Gray2005}. By embedding this evaluation directly into the likelihood step, this dataset self-consistently acts as the strongest anchor for $v \sin i_\star$ and $V_{\text{sys}}$, while simultaneously fitting for the intrinsic line depth and the normalisation continuum level.
\end{itemize}

\begin{table}[t] 
    \centering
    \caption{Mapping of the physical parameters constrained by each observational dataset within the joint Bayesian framework.}
    \renewcommand{\arraystretch}{1.3}
    \begin{tabular}{lcccc}
        \hline\hline
        Physical Parameter & TESS & RV & MAP & PROF \\
        \hline
        Core radius ratio, $R_p/R_\star$        & $\bullet$ &           &   $\bullet$        &           \\
        Planetary RV amplitude, $K_p$           &           &           & $\bullet$ &           \\
        Atmospheric wind, $v_w$                 &           &           & $\bullet$ &           \\
        Stellar RV amplitude, $K$               &           & $\bullet$ &           &           \\
        $(R_p/R_\star)_{\rm atmo}$              &           & $\bullet$ &           &           \\
        Orbital inclination, $i$                & $\bullet$ & $\bullet$ & $\bullet$ &           \\
        Projected obliquity, $\lambda$          &           & $\bullet$ & $\bullet$ &           \\
        RV offset, $\gamma$                     &           & $\bullet$ &           & $\bullet$ \\
        $v \sin i_\star$                        &           & $\bullet$ & $\bullet$ & $\bullet$ \\
        \hline
    \end{tabular}
    \tablefoot{The columns denote the different datasets modelled in our framework: TESS transit photometry (TESS), one-dimensional radial velocities (RV), two-dimensional tomographic map (MAP), and the out-of-transit 1D mean rotational profile (PROF).}
    \label{tab:parameter_matrix}
\end{table}

\begin{figure*}
    \centering
    \includegraphics[width=\linewidth]{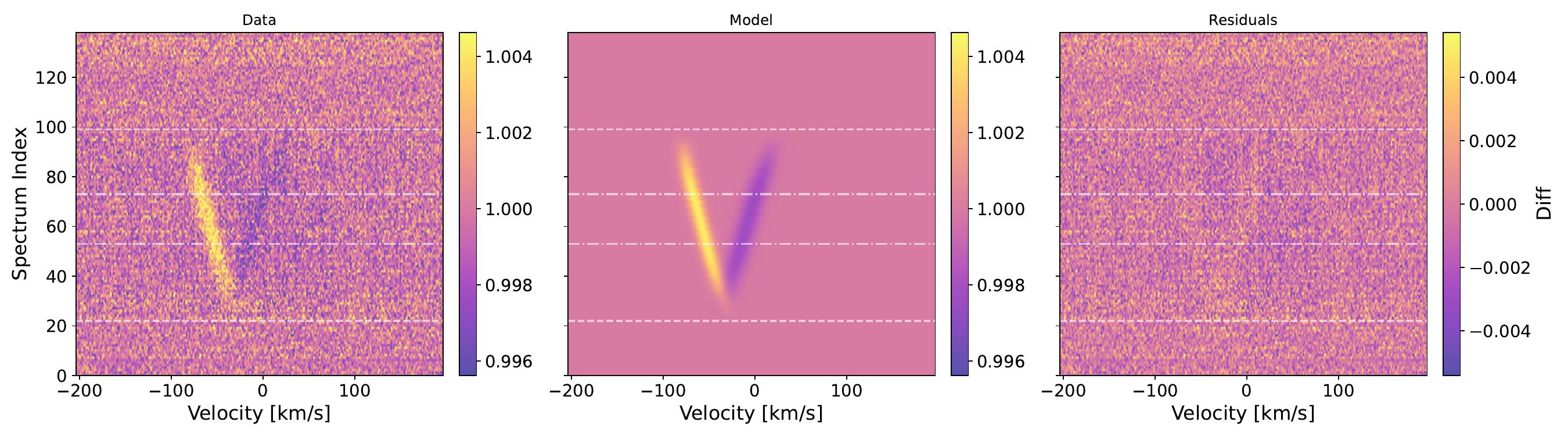}
    \caption{MAP results of the joint fit for TOI-1518 b. In the left panel is shown the original dataset, in the middle one the model and in the  right panel the residuals.}
    \label{fig:mcmc_map}
\end{figure*}

\begin{figure}
    \centering
    \includegraphics[width=\linewidth]{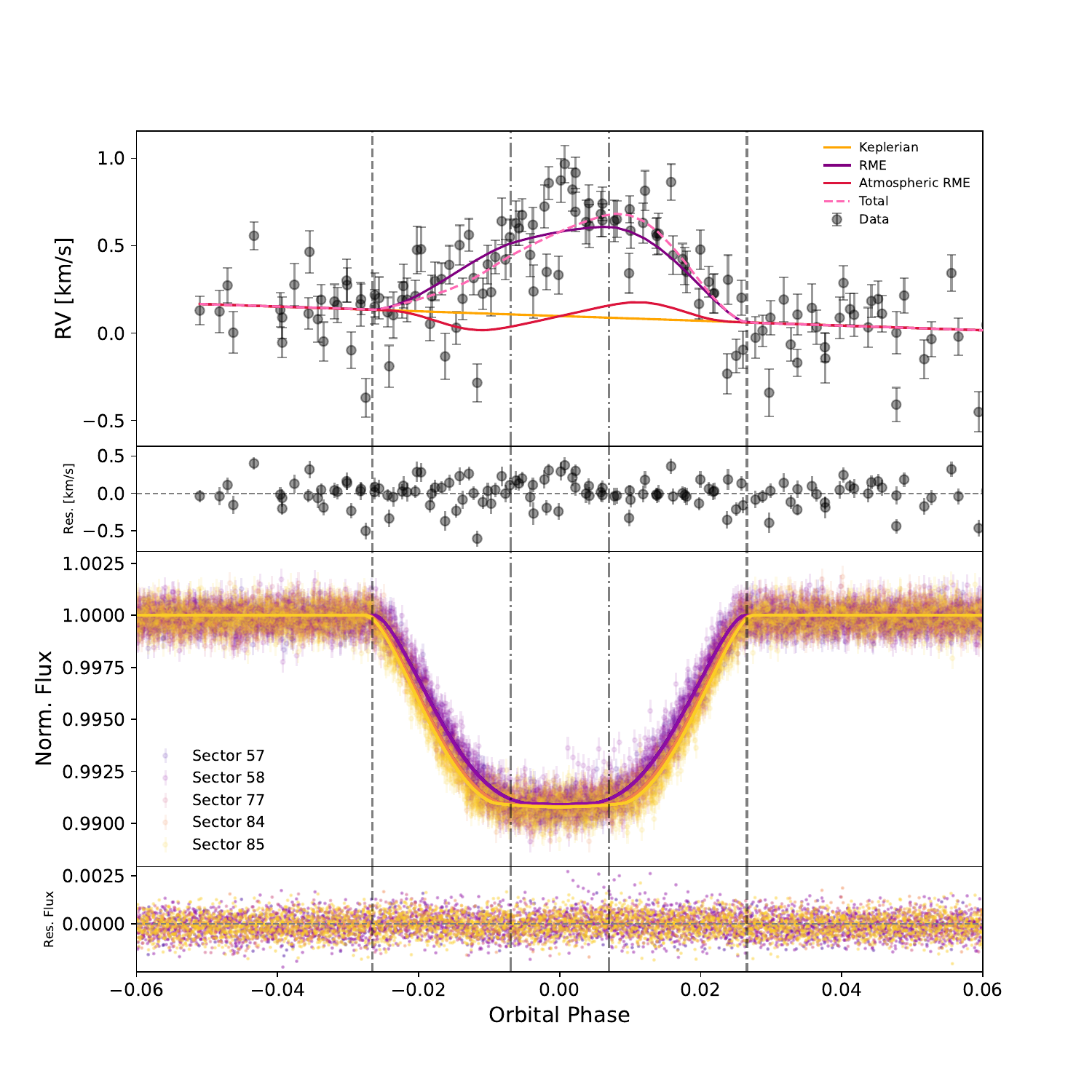}
    \caption{Comparison between the observational datasets and the global best-fit models as a function of the orbital phase for TOI-1518 b. \textit{Top:} HARPS-N radial velocities highlighting the RM effect. \textit{Bottom:} TESS photometry phase-folded around the mid-transit time. Both the photometric data and the theoretical models are colour-coded by TESS sector (57, 58, 77, 84, and 85) to visually highlight the transit shape variations induced by the measured nodal precession ($\mathrm{d}i/\mathrm{d}t$) across the observational baseline. Residuals are shown in the respective bottom panels.}
    \label{fig:mcmc_rv_lc}
\end{figure}

The results of our joint Bayesian fit, incorporating TESS photometry, the LSD-derived HARPS-N RVs, and tomographic LSD profiles, are summarized in Table~\ref{tab:toi1518_par}. The corner plot is shown in Fig.~\ref{fig:corner}, while the comparison between the datasets and theoretical models is shown in Figs.~\ref{fig:mcmc_map}, \ref{fig:mcmc_rv_lc}, and \ref{fig:prof}. Our analysis provides a refinement of the system parameters, which are generally in agreement with the discovery paper by \citet{cabot21}. One of the most notable improvements in our analysis is the determination of the stellar projected rotational velocity. We obtained $v \sin i_\star = 79.70^{+0.19}_{-0.47}$~\kms, a value significantly more precise than the $85.1 \pm 6.0$~\kms reported by \citet{cabot21}. This refinement is crucial for the accurate modelling of the Doppler shadow; by including the master out-of-transit LSD profile directly in the MCMC fit, we provided a direct spectroscopic constraint on the rotational broadening that is independent of the transit geometry. This lower and more constrained $v \sin i_\star$ has a direct impact on the derived orbital architecture. In tomographic modelling, the rotational velocity is strongly degenerate with the projected obliquity ($\lambda$) and the orbital inclination ($i_0$). We argue that our refined value, being anchored by the intrinsic stellar line morphology, effectively breaks this degeneracy. Consequently, the slight shifts in the orbital parameters observed in our work compared to previous studies are a direct result of this improved stellar characterisation. Regarding the system’s architecture, we measured a projected spin-orbit angle $\lambda = 238.21^{+0.43}_{-0.30}$~deg slightly offset from \citet{cabot21} ($\lambda = 240.34 \pm 0.32$~deg). Remarkably, by explicitly allowing the orbital inclination to vary over time across the multi-year TESS baseline, our joint fit achieved a significant empirical detection of orbital nodal precession. We measured an inclination change rate of $\mathrm{d}i/\mathrm{d}t = 0.1530 ^{+0.0036}_{-0.0035}$ deg yr$^{-1}$. Translating our measured inclination change into a variation of the transit impact parameter, we obtain $\mathrm{d}b/\mathrm{d}t = -0.01047 ^{+0.00063}_{-0.00062}$ yr$^{-1}$. This is in agreement with the value of $\mathrm{d}b/\mathrm{d}t = -0.0116 \pm 0.0036$\,yr$^{-1}$ recently reported by \citet{Watanabe2024}, providing a strong, independent confirmation of the orbital nodal precession using a comprehensive global framework. A key result of our joint fit is the detection of the atmospheric RM effect. We retrieved an effective atmospheric radius extension $(R_p/R_\star)_{\rm atmo} = 0.0794^{+0.0084}_{-0.0091}$. This represents a detection significance of approximately $8.5\sigma$, highlighting the impact of the planetary atmosphere on the stellar velocity field during the transit. This measurement, coupled with the core radius ratio $R_p/R_\star = 0.09892^{+0.00033}_{-0.00033}$, provides a robust baseline for the single-line transmission spectroscopy discussed in the following sections. The tomographic analysis (MAP) successfully isolated the Doppler shadow and the planetary signal, characterised by Gaussian amplitudes $A_1 = 0.004251 \pm 0.00009$ and $A_2 = 0.00225 \pm 0.00007$, respectively. The retrieval of a net atmospheric wind shift $v_{\rm w} = -7.02^{+0.42}_{-0.42}$~\kms suggests the presence of a strong day-to-night side transport or global circulation patterns in the atmosphere of TOI-1518~b, consistent with the high irradiation levels of the system.

\subsubsection{Mass refinements}
\begin{figure}
    \centering
    \includegraphics[width=\linewidth]{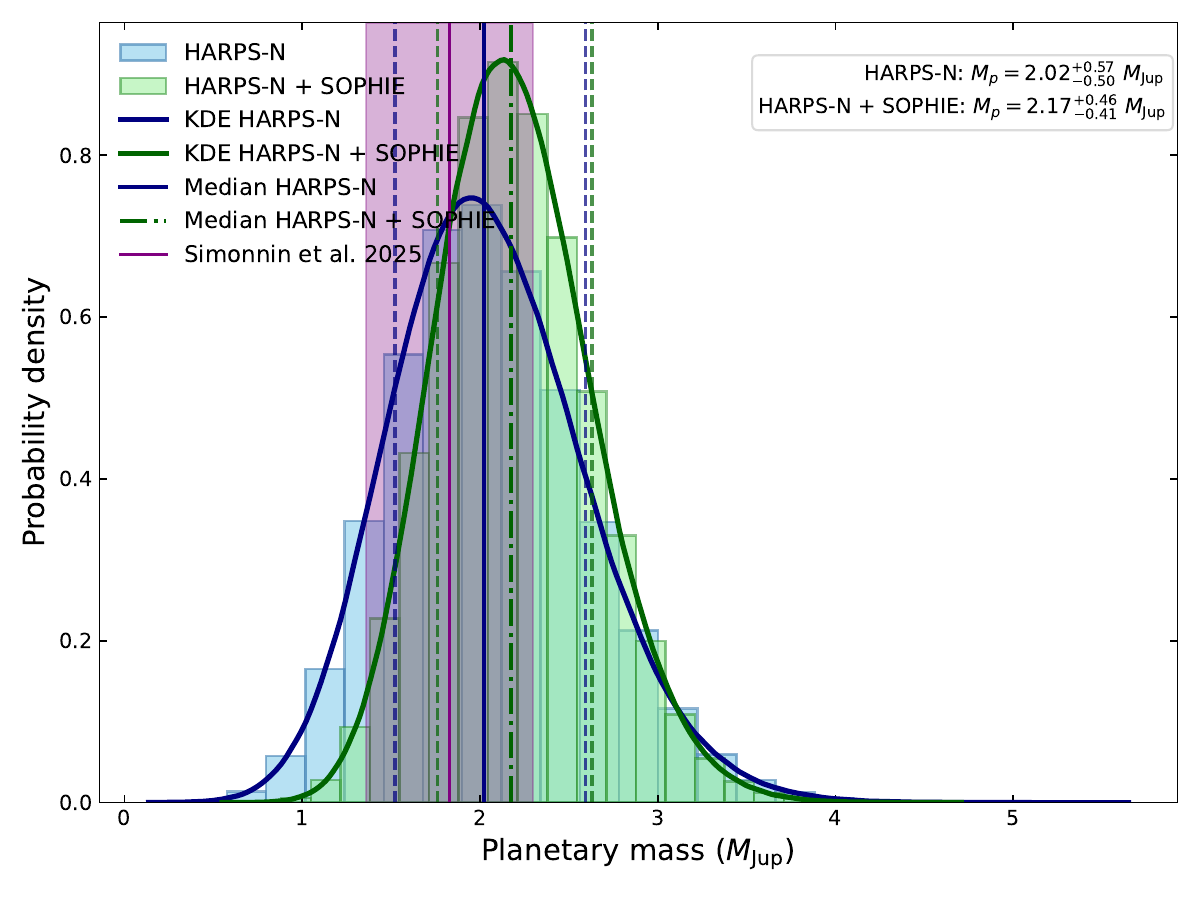}
    \caption{Planetary mass distribution obtained from our primary analysis of the HARPS-N data only, and for the one including the SOPHIE RVs. The magenta area represents the measurements of \cite{Simonnin2025}.}
    \label{fig:mass_both}
\end{figure}

The joint modelling of the RV curve and the tomographic signals (both planetary track and RM effect) allows for a direct measurement of the stellar reflex motion semi-amplitude, $K$, and the $K_p$. From our primary analysis using the HARPS-N dataset combined with TESS photometry and LSD profiles, we derived $K = 206^{+42}_{-43}$~\ms\ and $K_p = 188.1^{+6.6}_{-6.6}$~\kms, which is consistent within $1\sigma$ with the values reported in literature, confirming the nature of TOI-1518~b as a typical UHJ. Assuming a stellar mass $M_\star = 1.79 \pm 0.26\,M_\odot$ \citep{cabot21}, we calculated the planetary mass distribution through the mass function relation:
\begin{equation}
    M_p = M_\star \frac{K}{K_p},
\end{equation}
valid in the $M_p \ll M_\star$ approximation.
This leads to a planetary mass estimate of $M_p = 2.02^{+0.57}_{-0.50}$ \mjup, which is consistent within 1$\sigma$ with the recent literature values (e.g., $1.83 \pm 0.47$ \mjup; \citealt{Simonnin2025}). We show the planetary mass distribution in Fig.~\ref{fig:mass_both}. To further improve the precision of our mass estimate, we performed a secondary joint fit by incorporating the out-of-transit SOPHIE radial velocities reported by \citet{cabot21}, thereby extending the observational baseline. The resulting posterior distributions (see the corner plot in Fig.~\ref{fig:corner}) are centred on values identical to those of our primary analysis, demonstrating the robustness of our framework. Moreover, the inclusion of the SOPHIE data led to a significant reduction in the uncertainties. We obtained $M_p = 2.17^{+0.46}_{-0.41}$ \mjup, reducing the error on the stellar semi-amplitude $K$ and substantially improving the precision of the derived planetary mass $M_p$ as shown in Fig.~\ref{fig:mass_both}. The full-orbit RVs are displayed in Fig.~\ref{fig:full_rv}. This highlights the critical importance of achieving comprehensive orbital phase coverage, rather than relying solely on observations restricted to the transit window, to accurately sample the velocity peaks of the Keplerian curve. Capturing these quadrature points is essential to break parameter degeneracies, thereby significantly tightening the constraints on the mass and bulk physical properties of UHJs.

\subsection{Atmospheric characterisation}
\label{sec:atmo}
The extreme $T_{\text{eq}}$ and intense irradiation levels of TOI-1518~b make it a benchmark target for HRTS. The atmospheres of UHJs are expected to be dominated by atomic and ionic species, as most molecular bonds are dissociated by the high incident flux and by the high temperature. Notably, an indication of the planetary atmosphere is already discernible in the residuals of the stellar LSD profiles (Sect.~\ref{sec:results}), where a faint secondary trace following the planetary orbital motion emerges alongside the Doppler shadow, as can be seen in Fig. \ref{fig:mcmc_map}. 
We used two different methods to search for atmospheric signals, which are described in detail in the following sections. First, we analysed strong individual lines, such as the Balmer series. This approach is useful for studying the shape of the lines, which provides information on atmospheric depth and rotation. However, this method requires a very high S/N. To detect more species, we also used a CCF analysis with theoretical templates. By combining thousands of weak lines, the CCF is the best way to identify metals like \FeI\ and \FeII. Together, these two methods allow us to study both the chemical makeup and the wind patterns of TOI-1518~b.

\subsubsection{Single line analysis}
\label{subsec:single}
To extract the high-resolution transmission spectra, we followed the comprehensive framework thoroughly detailed in \cite{darpa_kelt9} while here we briefly highlight the key steps. After the standard data reduction and extraction using the 3.7 version HARPS-N DRS, we corrected the s1d spectra for telluric contamination using the line-by-line radiative transfer model {\tt\string Molecfit} \citep{Smette_2015,Kausch_2015}. The telluric-corrected spectra were then normalised and shifted to the stellar rest frame using the systemic velocity ($V_{\mathrm{sys}}$) to construct 2D tomographic maps. A critical step in our analysis is the accurate removal of the RM effect and centre-to-limb variations (CLV), which heavily contaminate the 2D maps of fast-rotating stars. As described in \cite{darpa_kelt9}, we modelled these effects by computing the rotationally broadened spectrum of the dynamically obscured stellar disk at each orbital phase, relying on ATLAS9 stellar models \citep{kurucz1992model} and an analytical quadratic limb darkening approach. Once the RM effect and CLV contaminations were successfully isolated and removed from the stellar rest frame spectra, we shifted all the in-transit frames to the planetary rest frame according to the $K_{\text{p}}$. The final 1D transmission spectra for the individual lines were then obtained by computing the error-weighted average of these shifted in-transit spectra.

\subsubsection{Cross-correlation with templates}

Along with the single line analysis, we also performed a cross-correlation with templates employing the method described in detail in \cite{Biassoni24} and briefly explained in this section.
For our CCF analysis, we used the HARPS-N echelle spectral orders e2ds \footnote{In the HARPS-N Data Reduction Software, e2ds files contain the extracted two-dimensional echelle spectrum (one row per spectral order), while s1d files contain the one-dimensional spectrum obtained by merging all orders and resampling them onto a uniform wavelength grid.}. Telluric absorption from H$_2$O and O$_2$ was corrected by interpolating the {\tt\string Molecfit} transmission spectrum (derived from the s1d spectra, as detailed in Sect. \ref{subsec:single}) onto the e2ds wavelength grid. We restricted the final wavelength range to $\sim$4001--6832 \AA{} to ensure optimal signal-to-noise for the CCF extraction. Specifically, we excluded the bluest five orders due to dropping instrumental throughput, and the reddest order because of decreasing detector efficiency combined with extremely dense telluric contamination.

Prior to performing the CCF, the e2ds echelle orders were normalised and stacked.
The outcome is a set of one-dimensional spectra, $x(t)$, one for each exposure, $t$, structurally similar to the HARPS-N s1d spectra but benefiting from improved continuum normalisation and preserving the original wavelengths of the echelle orders.

These reconstructed spectra $x(t)$ were then used as input for the CCF analysis, which was carried out independently for each night. We adopted the normalised CCF definition:

\begin{equation}\label{CCF_norm}
CCF(v,t) = \frac{\sum_{k}{x_k(t) \cdot T_k(v)}}{\sum_k T_k(v)} \,,
\end{equation}

\noindent
where $v$ represents the velocity shift applied to the template $T$ of a species, and $k$ indexes the pixels of the reconstructed 1D spectrum $x(t)$.
The templates were shifted over a velocity range of $-200$\,\kms \, $\leq v \leq +200$\,\kms \, in 1\,\kms \, increments and interpolated along the corresponding spectrum axis.
The resulting CCFs were then Doppler shifted into the stellar rest frame interpolating the RV within the range of $-150$\,\kms \, $\leq v \leq +150$\,\kms \,, also in steps of 1\,\kms \, and subsequently normalised by the temporal mean of the out-of-transit CCFs.
To further refine the normalisation, each CCF map was divided by the median across all exposures.
This procedure was applied independently for every inspected elements, temperatures and for each night of observations.

Although the actual atmospheric structure of TOI-1518b is expected to be complex \citep[e.g.,][]{Simonnin2025}, the high-resolution cross-correlation technique primarily responds to the relative positions and overall pattern of absorption lines rather than to the precise continuum level or the detailed vertical temperature profile. In this context, isothermal templates can still serve as effective matched filters for detecting specific species.
To account for possible variations in the atmospheric temperature, we performed the CCF analysis using a set of standard isothermal templates at 2000\,K, 2500\,K, 3000\,K, 4000\,K, and 5000\,K taken from \cite{2023A&A...669A.113K} and we convolved them to the instrumental resolution of HARPS-N.
This approach ensures that the templates adequately sample a range of atmospheric conditions and maximises the likelihood of matching the absorption lines of different chemical species, allowing us to detect elements that might be missed if only one or two template temperatures were used \citep{Biassoni_2026}.

\paragraph{Doppler shadow correction}
\label{Doppler shadow correction}

During a planetary transit, the stellar absorption lines are distorted due to the portion of the disc occulted by the planet, producing a time-dependent feature in the CCF maps known as the Doppler shadow. The shape and amplitude of the Doppler shadow depend on the projected stellar rotational velocity, the projected spin-orbit angle, the impact parameter, the $R_p/R_{\star}$ ratio, and also on the specific element being analysed.

We modelled the radial velocities of the Doppler shadow ($v_{\mathrm{DS}}$) following \cite{Rainer_2021}, which can be expressed as:
\begin{equation}\label{eq:vds}
v_{\mathrm{DS}} = v_{\star} \sin{i}\,  (x_p \cos{\lambda} - y_p \sin{\lambda}) \,,
\end{equation}

\noindent
where $x_p = a_{R_{\star}} \sin{(2 \pi \phi)}$ and $y_p = - a_{R_{\star}} \cos{(2 \pi \phi)} \cos{i}$, with $a_{R_{\star}}$ being the semi-major axis in units of the stellar radius, $\lambda$ the projected obliquity, $i$ the orbital inclination, $v_{\star} \sin{i}$ the projected rotation speed, and $\phi$ the orbital phase.

We corrected each element and each night individually by fitting a double Gaussian to each exposure (i.e. each orbital phase in the CCF), centred at the Doppler shadow velocity predicted by Eq.~\ref{eq:vds} \citep[e.g.,][]{Hoeijmakers_2020,Rainer_2021}. Correcting for the Doppler shadow is essential to avoid misinterpreting planetary signals in the CCF maps.
The correction was applied only when the Doppler shadow could be reliably fitted. If the fit was not statistically significant, the Doppler shadow was left uncorrected, thus preventing the introduction of artificial noise into the data.

\paragraph{Merged CCF} \label{Merged CCF}

After obtaining the CCFs for each element and night and correcting for the Doppler shadow, we pooled the CCF profiles from all nights and arranged them in ascending order of their orbital phases. This resulted in a single consolidated CCF map for each element and template temperature where each row represents a single phase observation and no weighting scheme was applied.
To improve the normalisation of the CCF maps, a Fourier filter was applied to remove velocities exceeding 200\,\kms . In addition, to mitigate residual stellar or telluric features, as well as other systematic noise that is constant over time, we fitted a first-degree polynomial to each column (RV) of the consolidated two-dimensional CCF and divided them by this fit \citep[e.g.][]{Prinoth_2022,Biassoni_2026}. Finally, the CCF continuum was set to zero by subtracting 1.
To enhance the planetary signal, we derived the $K_p$–$V_{\mathrm{sys}}$ maps \citep{Brogi_2012} for each CCF, scanning $K_p$ from 0 to 350\,\kms \, in steps of 1\,\kms. To express the maps in terms of S/N, we calculated the standard deviation $\Sigma$ excluding the RV range $[-35, +35]$\,\kms \, and normalised the maps by this value.
The analysis accounts for both the systemic velocity and the barycentric Earth RV, so the planetary signal is expected to be centred at RV$=0$\,\kms.
Given the large number of elements analysed, we defined a clear criterion to distinguish detections from non-detections. In our $K_p$–$V_{\mathrm{sys}}$ maps, we consider a signal as a detection if it exhibits S/N $\geq 4$ within a $K_p$ range of 100–250\,\kms \, and within $\pm 10$\,\kms \, of RV $= 0$\,\kms. Signals with $3.5 \leq \mathrm{S/N} < 4.0$ under the same $K_p$ and RV constraints are classified as tentative detections as in \cite{Biassoni_2026}. While these thresholds are routinely adopted in HRTS studies, we note that their statistical robustness against false positives is significantly reinforced in our analysis by coupling them with two strict criteria. First, any candidate signal must strictly match the expected physical constraints we adopted in the $K_p$–$V_{\mathrm{sys}}$ parameter space. Second, it must be validated by our aliasing analysis to rule out periodic artifacts or noise-driven fluctuations. Therefore, the combination of a high S/N peak, strict kinematic alignment, and the absence of strong aliases ensures that our claims are statistically secure.

It is possible that some of the signals fulfilling our detection criteria may arise from aliasing rather than from true planetary absorption \citep{borsato2023mantis}. Aliasing occurs when spectral lines of different templates partially overlap, producing spurious signals in the CCF and $K_p$–$V_{\mathrm{sys}}$ maps that can mimic a real planetary signature, particularly when templates contain only a limited number of weak lines.
To identify and correct these effects, we adopted the same strategy described in detail in \cite{Biassoni_2026}.
Basically, for each detected or tentatively detected species at a given temperature, we generated synthetic CCFs by cross-correlating the template of the inspected element with Doppler-shifted templates of all other detected species, assuming the $K_p$ value corresponding to the minimum of the observed $K_p$–$V_{\mathrm{sys}}$ map. This procedure allowed us to evaluate whether the observed signal could be reproduced by cross-correlating different templates. Synthetic aliasing signals that exceeded the noise level of the observed maps were treated as potential contaminants. These contributions were modelled through a multiple regression approach with Tikhonov regularization and subtracted from the observed CCFs, yielding corrected maps for each species.
When necessary, Doppler shadow features associated with aliasing signals were removed prior to the regression analysis, following the same prescription described in Sect.~\ref{Doppler shadow correction}.
This procedure ensures that only statistically robust and physically consistent signals are retained as genuine atmospheric detections.

\subsubsection{Results}
\label{sec:results}
\begin{figure*}
    \centering
    \includegraphics[width=0.3\linewidth]{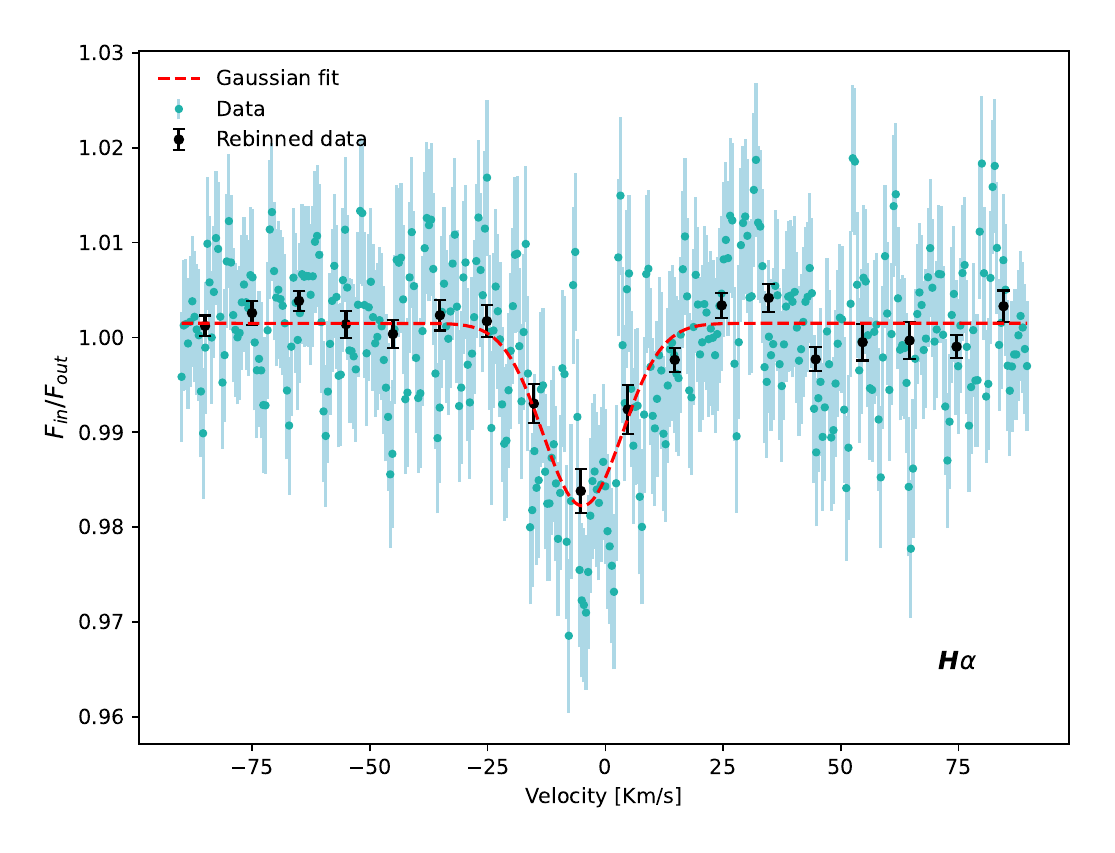}\quad
    \includegraphics[width=0.3\linewidth]{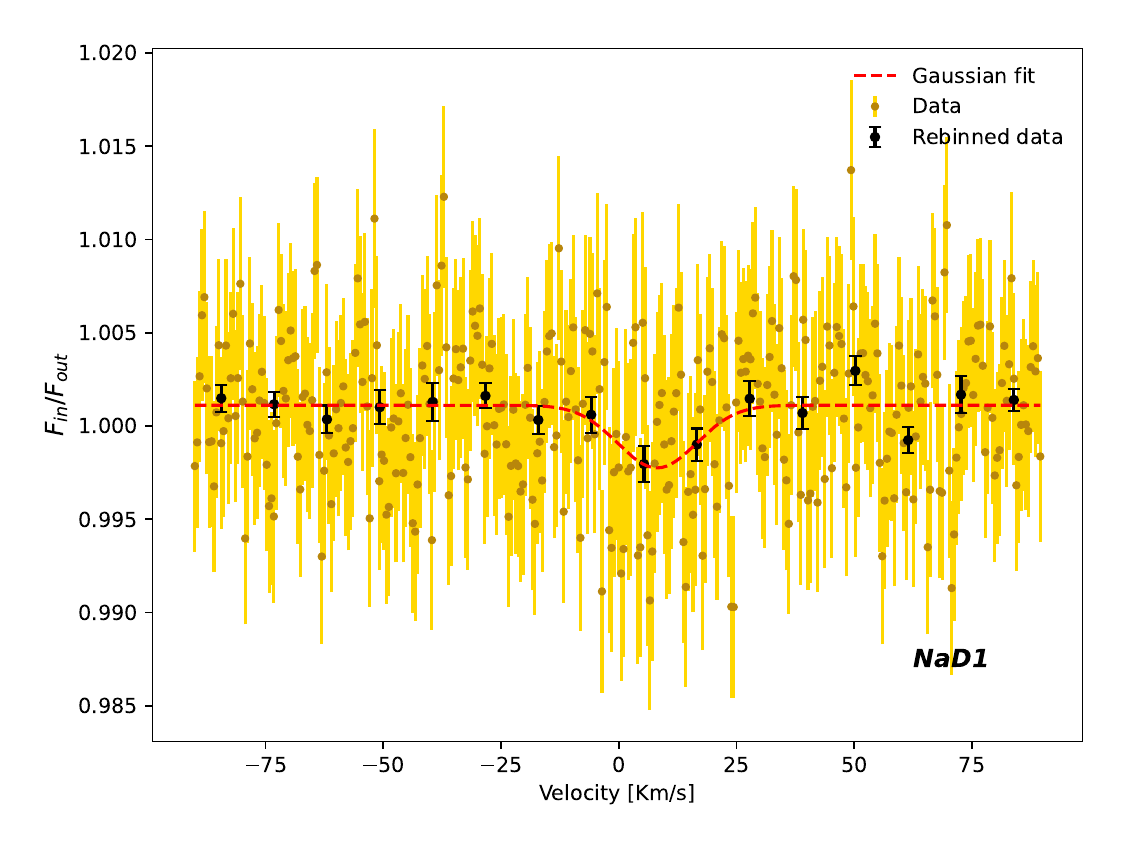}
    \includegraphics[width=0.3\linewidth]{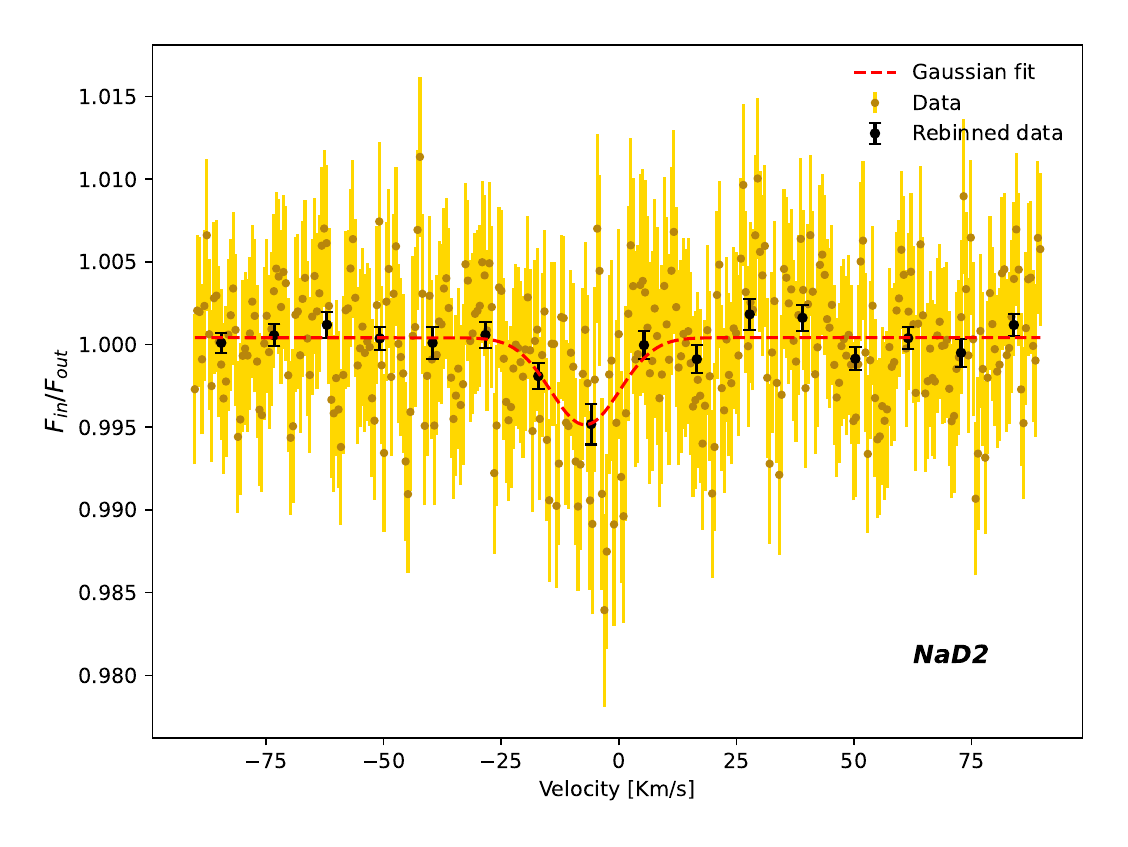}\\
    \includegraphics[width=0.3\linewidth]{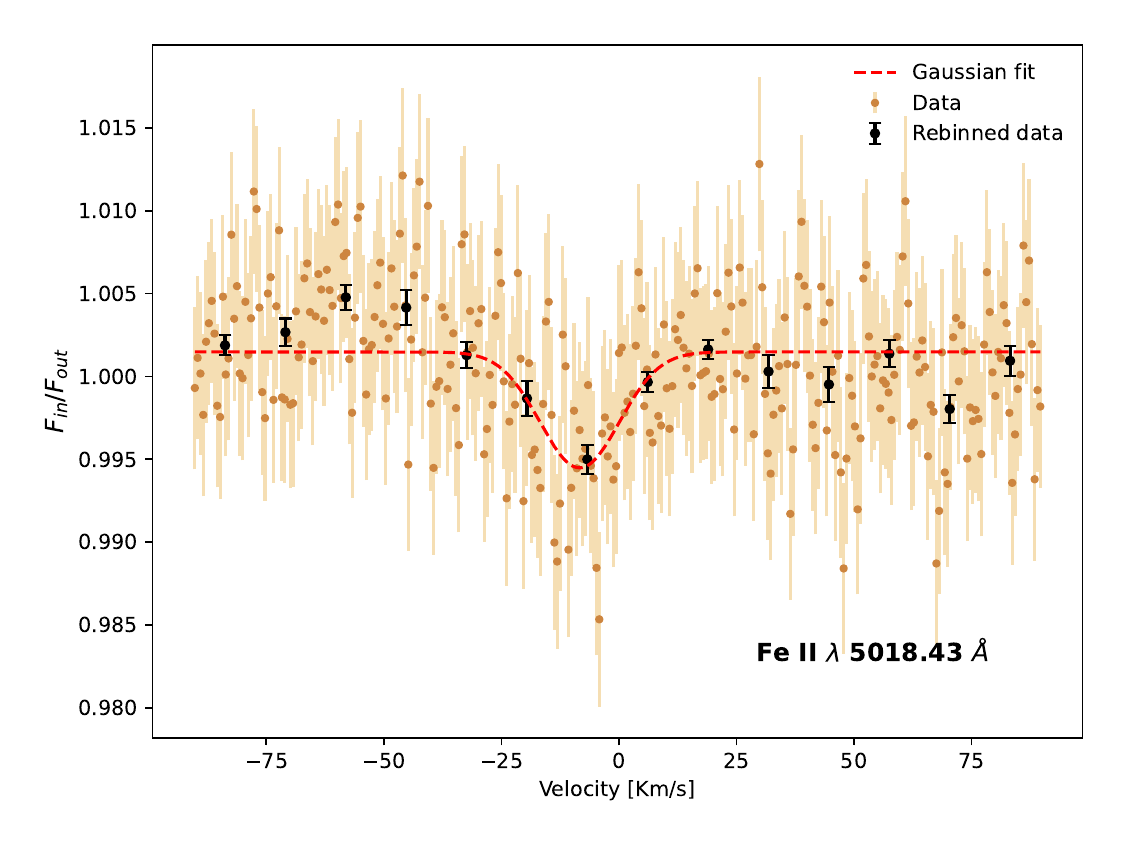}
    \includegraphics[width=0.3\linewidth]{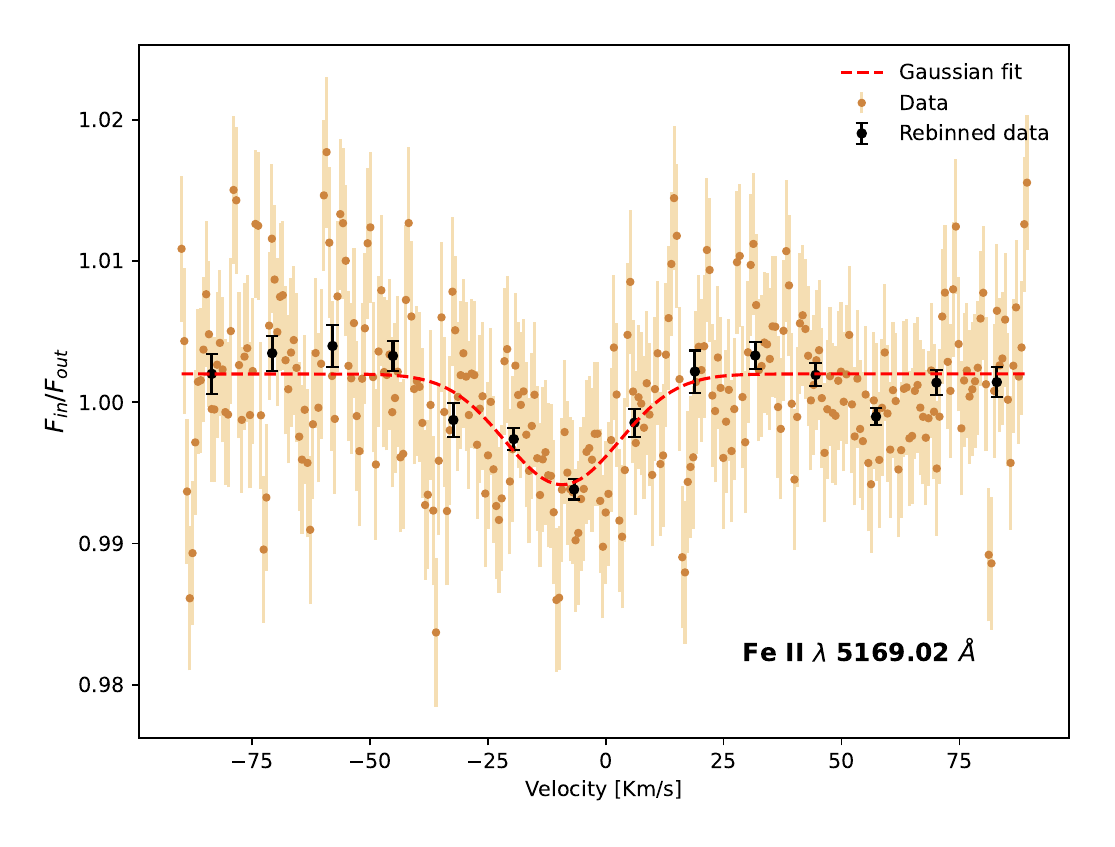}\\

    \caption{One-dimensional transmission spectra for individual atomic and ionic species detected or investigated in the atmosphere of TOI-1518~b, extracted in the planetary rest frame. We show $\text{H}\alpha$ (top), the \NaI\,D1 and D2 doublet (middle), and two prominent \FeII\, transitions at 5018.43~\AA\ and 5169.02~\AA\ (bottom). In all panels, the light-coloured dots represent the individual data points, the black squares with error bars denote the rebinned data, and the dashed red lines show the best-fit Gaussian profiles.}
    \label{fig:singles_opt}
\end{figure*}

We individually analysed several lines belonging to \HI, \NaI, \CaII, \TiII, \FeI, \FeII, \BaII, \MgI, \CrII and \ScII analogously to what we did for benchmark planet KELT-9 b \citep{darpa_kelt9}, showing the detections in Fig. \ref{fig:singles_opt}. Among the hydrogen Balmer series, we successfully detected the \Ha\ line, retrieving an absorption depth of $1.93 \pm 0.17\%$ and a blue-shift of $v_{\rm wind} = -4.84 \pm 0.84$~\kms. No significant absorption was found for the higher-order transitions, which is consistent with the lower S/N and the intrinsically weaker nature of these lines. To date, these have only been robustly detected in a small subset of the most extreme targets, such as HD 189733\,b, KELT-9\,b, and KELT-20\,b \citep{cauleyhd, 2019A&A...628A...9C}. Apart from the lines listed in Table~\ref{tab:single_line_results}, we were not able to detect any other individual line. We were also not able to investigate the Ca {\sc ii} H \& K lines due to the low S/N in the bluest part of the HARPS-N spectra. Regarding the other investigated species, while the \NaI\ doublet initially showed signals exceeding $3\sigma$, the physical inconsistencies in their velocity shifts (e.g., $+8.1 \pm 2.8$~\kms\ for the D1 line versus $-7.2 \pm 1.7$~\kms\ for the D2 line) suggest that these features are likely dominated by stellar residuals rather than planetary absorption, as further discussed in Sect.~\ref{sec:discussion}. Notably, our single-line search also revealed two prominent absorption features associated with \FeII\ at 5018.43~\AA\ and 5169.02~\AA. Both transitions yield consistent, high detection significances ($>6\sigma$) and physical parameters, displaying a blue-shift (averaging $\sim -8.8$~\kms) that agrees with the global atmospheric wind patterns derived from our joint analysis.

From the analysis of the CCFs, the Gaussian fit along the RVs gave a $> 4$ S/N detection of \HI \,, \CI \,, \NaI \,, \CaI \,, \ScII \,, \TiII \,, \VII \,, \CrI \,, \FeI \,, \FeII \,, \NiI \,, \SrII \,, \YII \,, \EuII \,, and \WI \,, while we tentatively detected \LiI \,, \MgI \,, \BaI \,, \BaII \,, and \NdII.
After carefully accounting for potential aliasing effects, we found that all detected signals can be considered robust. In particular, the aliasing analysis confirms that all the detections identified in the CCF and $K_p$–$V_{\mathrm{sys}}$ maps are genuine, and no correction for aliasing \citep{Biassoni_2026} is required. This validation step is especially important given the presence of several uncommon species that have not been previously reported.
We computed the merged (between nights) CCFs and corresponding $K_p$–$V_{\mathrm{sys}}$ maps for all species across the full grid of five temperatures considered. For each species that passed the detection threshold as indicated in Sect. \ref{Merged CCF}, we extracted the $K_p$ value that maximises the signal in its $K_p$–$V_{\mathrm{sys}}$ map. At this optimal $K_p$, we fitted the corresponding one-dimensional CCF with a Gaussian profile plus a constant offset along the RVs, deriving the centre, FWHM, and amplitude of the signal (similar to \citealt{Hoeijmakers2019} and \citealt{borsato2023mantis}).
Conversely, the 1D vertical profile along the $K_p$, taken at the RV of the peak signal, typically deviates from a Gaussian distribution. To avoid the systematic biases inherent in parametric fitting for such non-Gaussian shapes, we determined the value and its associated uncertainty using a bootstrap method rather than a standard fit as in \cite{Biassoni_2026}.

Given the large number of in-transit spectra, we further divided the transit into two temporal segments: from $T1$ to mid-transit (pre-$T_0$), and from mid-transit to $T4$ (post-$T_0$), inspecting in this way the morning and evening side of the planet. For each subset, we recomputed the $K_p$–$V_{\mathrm{sys}}$ maps, and derived the corresponding detection parameters using the same Gaussian fitting procedure along the RVs at the $K_p$ value that maximises the signal.
All our results are summarised in Fig. \ref{fig:kpvsys_fit} where we report, for each species and temperature, the S/N obtained from the Gaussian fit measured over the full transit as well as in the pre- and post-$T_0$ phases. The significance of the detection is indicated using marker sizes and colour coding to highlight detections versus tentative detections and non-detections. 
The empty regions represent cases where no coherent signal was identified above the pre-selection threshold in the $K_p$–$V_{\mathrm{sys}}$ maps, as stated in Sect.~\ref{Merged CCF}. In these instances, performing a parametric fit was considered both superfluous and statistically unreliable due to insufficient S/N.

It is possible to note how \CI \,, \MgI \,, \ScII \,, \VII \,, \CrI \,, \SrII \,, and \BaI \, exhibit a stronger signal during the second half of the transit (see also Table \ref{tab:detection}). The average S/N and number of spectra of our sample of four nights in the pre-$T_0$ phases are $55 \pm 11$ and 41, while post-$T_0$ phases we have $56 \pm 7$ with 36 total exposures. Hence, the highest detection observed in the post-$T_0$ phases does not appear to be driven by differences in the average S/N, number of spectra, or telluric residual, given the CCF cleaning procedure described in Sect. \ref{Merged CCF}. For these reasons, the limb differences in the S/N observed in different species, may be associated with a physical effect instead of noise or artifacts. It may be attributed to a higher temperature on the evening limb, as predicted by general circulation models of tidally locked planets, where stellar irradiation and zonal winds shift the hot spot eastward \citep{Showman_2011}, increasing the atmospheric scale height and producing a greater absorption depth along the line of sight. Interestingly, among these species, \MgI \,, \CrI \,, and \BaI \, have a lower first ionisation energy compared with \FeI \, that is detected on both the limb and is detected also in its ionised state. Therefore, one might expect these species to be ionised rather than neutral on the day side due to its much higher temperature compared to the night side. Their disappearance specifically during the morning limb suggests a heterogeneous distribution or cold trap mechanism \citep[e.g.][]{Parmentier_2013,Parmentier_2016,Hoeijmakers_2024,Pelletier_2026}.

\begin{figure}
    \centering
    \includegraphics[width=\linewidth]{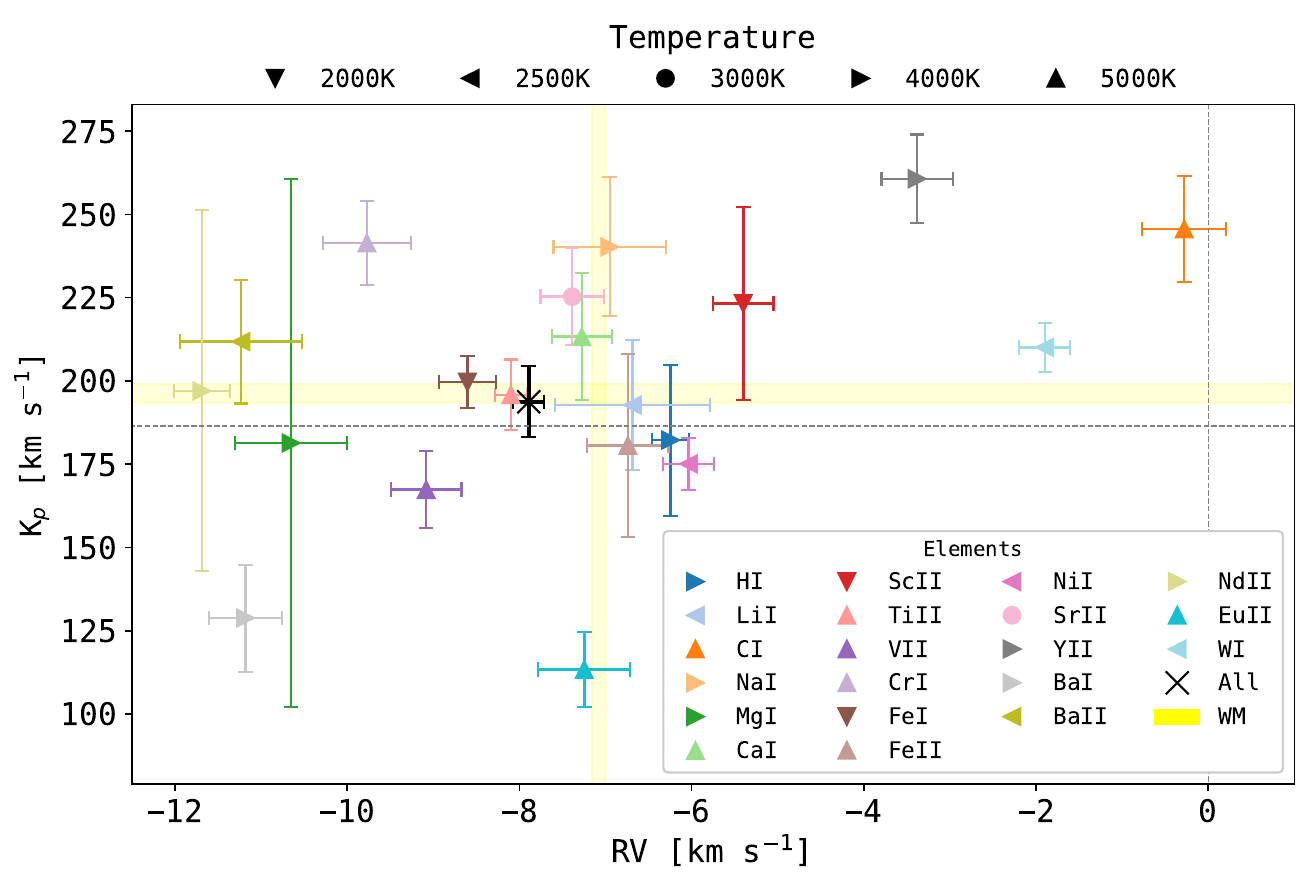}
    \caption{RV and $K_p$ values of detected species. For each element, we show the RV derived from the Gaussian $+$ offset fit and the $K_p$ obtained from the bootstrap. Only the detection at the temperature that maximises the S/N is shown for each species. Vertical and horizontal dashed gray lines indicate the reference frame (i.e., RV = 0\,\kms \,) and the theoretical planetary semi-amplitude $K_p$, while the yellow bands are the weighted mean of RVs and $K_p$.}
    \label{fig:KpRV}
\end{figure}

\begin{figure*}
    \centering
    \includegraphics[width=\linewidth]{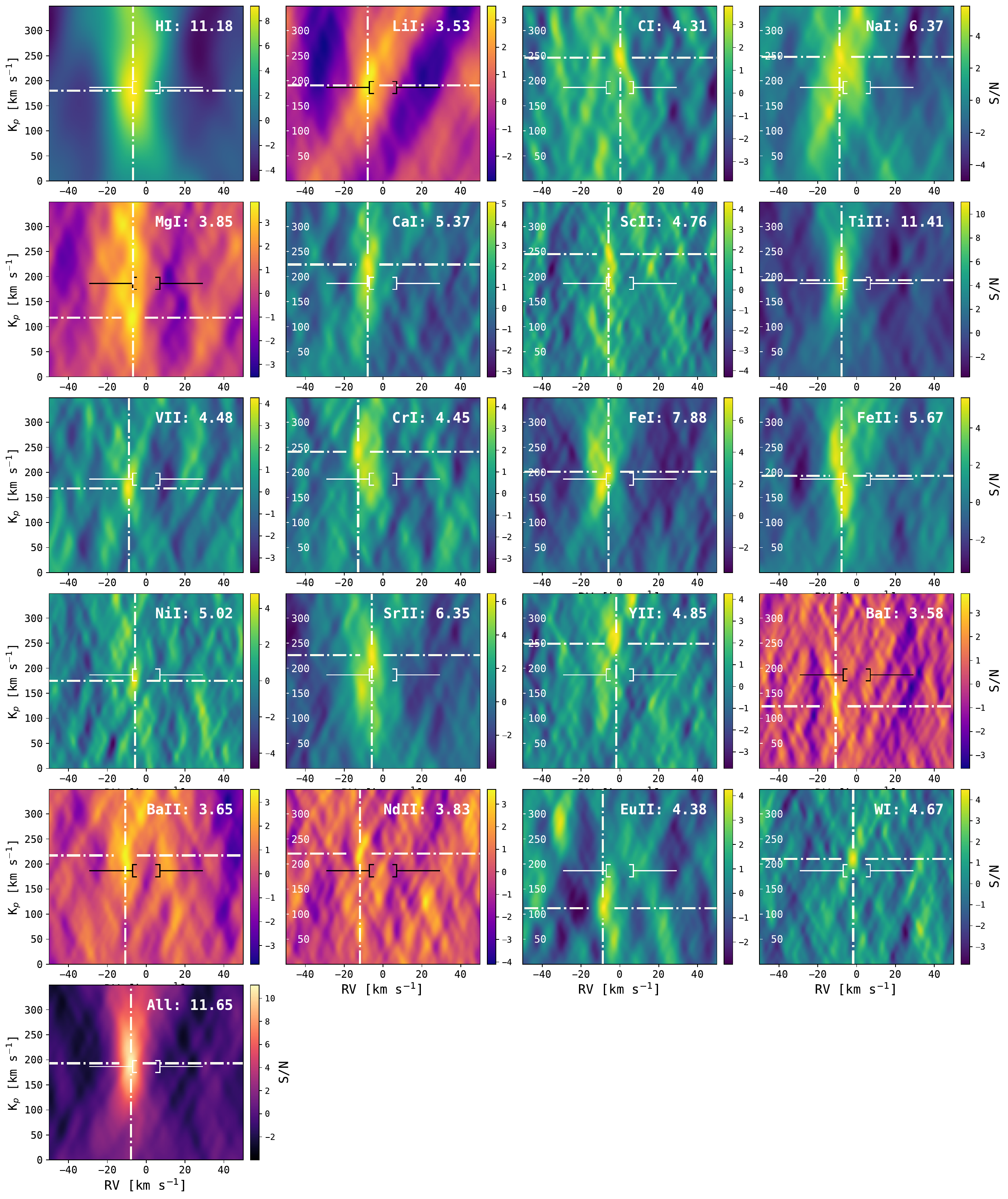}
    \caption{$K_p$–$V_{\mathrm{sys}}$ maps for all detected and tentatively detected species, shown at the temperature that maximises the S/N. Violet-yellow panels indicate tentative detections. The white and black sight mark the expected planetary signal, corresponding to RV = 0\,\kms \ and $K_p$ = 188.1\,\kms. The ivory lines highlight the location of the maximum signal in each map. The S/N values quoted in the text are derived from Gaussian fits to the corresponding one-dimensional $K_p$–$V_{\mathrm{sys}}$ profiles, evaluated at $K_p^{\rm max}$.}
    \label{fig: ccf bias}
\end{figure*}

Finally, we summarise the main results in Table \ref{tab:detection}, where we report only the detections corresponding to the temperature that maximises the full-transit S/N for each species. For consistency, the pre- and post-$T_0$ S/N values are evaluated at the same temperature, even when higher S/N values may be obtained at different temperatures in the individual subsets. The table also includes the fitted RV centroids obtained from the Gaussian fit and the $K_p$ values with their associated uncertainties derived from the bootstrap analysis.
Figure \ref{fig:KpRV} shows all the detected species at the temperature that maximises their S/N with the $K_p$ and RV values from Table \ref{tab:detection}. The weighted mean centre RV of all such species is at $-7.084 \pm 0.079$\,\kms \, showing a day-to-night side wind, while the mean $K_p$ value is at $196.2 \pm 2.9$\,\kms \, slightly above the theoretical one, both highlighted with a yellow band. The associated $K_p$–$V_{\mathrm{sys}}$ maps are depicted in Fig. \ref{fig: ccf bias}.

We constructed a composite template by selecting, for each species, the template corresponding to the temperature that maximises its S/N. These templates were then summed to obtain a single combined template including all detected species \citep[e.g.,][]{2025A&A...694A.284P}. Using this composite template, we computed the CCF for each observing night, applied the Doppler shadow correction, and compute the final CCF by merging all the CCFs from each night and sorting them along the phases. From this CCF, we derived the corresponding $K_p$–$V_{\mathrm{sys}}$ map presented in the last panel of Fig. \ref{fig: ccf bias}. We fitted the Gaussian plus constant offset profile along the maximum $K_p$ value found in this $K_p$–$V_{\mathrm{sys}}$ map to extract its RV centre, while for the $K_p$ we derived it using the bootstrap. The fit results for the full transit, pre- and post-$T_0$ are summarised in Table \ref{tab:detection}, while the RV and $K_p$ values are plotted in Fig. \ref{fig:KpRV}.

\section{Discussion}
\label{sec:discussion}
In this study, we provided a comprehensive re-evaluation of the TOI-1518 system, bridging the gap between its dynamical architecture and its atmospheric composition. By leveraging a joint Bayesian framework, we established a self-consistent link between the system's complex orbital geometry and the physical properties of its ultra-hot atmosphere. By anchoring the $v \sin i_\star$ through the master out-of-transit LSD profiles, we broke the geometrical degeneracies typical of fast-rotating stars. This allowed us to confirm the polar architecture of the system with unprecedented precision, while simultaneously revealing a global atmospheric blue-shift ($v_{\rm w} = -7.02^{+0.42}_{-0.42}$~\kms) from our joint modelling. We note that the exact magnitude of this wind is highly sensitive to the adopted systemic velocity ($V_{\mathrm{sys}}$). While our HARPS-N dataset yields $v_{\rm sys} = -10.103 \pm 0.074$~\kms, literature values exhibit significant scatter across different instruments and epochs, ranging from $-10.22 \pm 0.35$~\kms\ (\textit{Gaia} DR3) to values between $-11.1$ and $-14.8$~\kms\ \citep{cabot21, Watanabe2024, Petz2025, Simonnin2025}. Adopting a more negative $V_{\mathrm{sys}}$ would proportionally decrease the magnitude of our retrieved $v_{\rm w}$. Nevertheless, across all plausible $V_{\mathrm{sys}}$ scenarios, the atmospheric signal remains strictly blue-shifted, consistently pointing to the presence of strong day-to-night side winds at the terminator.

Crucially, our time-dependent transit modelling yielded a significant empirical detection of orbital nodal precession ($\mathrm{d}i/\mathrm{d}t = 0.1530^{+0.0036}_{-0.0035}$ deg yr$^{-1}$) providing a robust observational validation of the theoretical framework proposed by \citet{Watanabe2024}, confirming the extreme dynamical environment shaped by the oblateness of the rapidly rotating host star.

We investigated the presence of the sodium doublet (\NaI\ D1 and D2) in the planetary transmission spectrum, finding absorption signals that exceed the $3\sigma$ threshold. However, a detailed comparison of their velocity shifts reveals a significant physical inconsistency. While the \NaI\ D2 line displays a blue-shift of $-7.2 \pm 1.7$~\kms, consistent with the expected atmospheric wind patterns, the \NaI\ D1 line appears strongly red-shifted at $+8.1 \pm 2.8$~\kms, suggesting that the Na features are likely contaminated by spurious stellar residuals. The contamination is specific to the \NaI\ D1 and D2 transitions as these resonance lines are deep and broad. As a result, even minor imperfections in modelling the stellar spectrum, such as residual CLV or 1D RM effect mitigation, can leave significant stellar residuals that distort the single-line profiles. In contrast, heavy metals and trace species (e.g., \WI, \EuII, \YII) exhibit intrinsically weak or non-existent photospheric absorption lines at these high stellar temperatures. Consequently, we expect their extraction to be immune to macroscopic stellar residuals. The robustness of these weaker detections is further guaranteed by the CCF technique, which combines all the lines of a given species included in the template, effectively averaging out localised spectral systematics.
In contrast, our single-line search for ionised iron lines identified two prominent features at 5018.43~\AA\ and 5169.02~\AA, representing the strongest expected transitions. These lines show a remarkable consistency in their retrieved parameters, with a clear blue-shift (averaging $\sim -8.8$~\kms) and a shared effective radius $R_p/R_\star \approx 1.34$. This robust individual detection is powerfully expanded by the full CCF analysis across all species. As depicted in the $K_p$--$V_{\mathrm{sys}}$ maps in Fig.~\ref{fig: ccf bias}, the weighted mean centre RV of all detected species is $-7.084 \pm 0.079$~\kms, matching our global MCMC wind parameter and acting as a clear indicator of a strong day-to-night side flow. We note that this global wind represents an integrated bulk metric of the upper atmosphere. As illustrated by the velocity dispersion of the individual species in Fig.~\ref{fig: ccf bias}, the atmosphere remains highly stratified. The variations in the retrieved RVs for different elements naturally reflect vertical wind shear and species-specific kinematics at different altitudes, consistent with recent 3D predictions and observations of similar UHJs \citep[e.g.,][]{2024PASP..136h4403W, 2026NatAs..10.1145S}. Simultaneously, the mean planetary semi-amplitude is retrieved at $K_p = 196.2 \pm 2.9$~\kms (highlighted with a yellow band in the maps). Although slightly above the theoretical Keplerian value ($K_p \approx 188$~\kms), this fully validates the planetary origin of the signals. The agreement between the single-line morphology of the strongest transitions and the global statistical significance of the CCF provides compelling evidence for the presence of heavy metals and ionised species in the upper atmosphere of TOI-1518~b. While other species were probed, they remained unidentified in individual lines due to local S/N limitations or line dilution, further highlighting the necessity of the CCF approach for comprehensively assessing the atmospheric inventory.

From our CCF and aliasing analysis, we were able to detect \HI,  \CI,  \NaI,  \CaI,  \ScII,  \TiII,  \VII,  \CrI,  \FeI,  \FeII,  \NiI,  \SrII,  \YII,  \EuII,  and \WI. Moreover, we tentatively detected \LiI,  \MgI,  \BaI,  \BaII,  and \NdII. While previous high-resolution campaigns successfully identified atmospheric species such as \FeI, \FeII, \NaI, \CaI, \CrI, \NiI, and \MgI \ \citep[e.g.,][]{cabot21, Simonnin2025, Basinger2025}, our comprehensive analysis and self-consistent RM effect mitigation allow us to report the first robust detection of \HI, \CI, \ScII, \VII, \SrII, \YII, \EuII, and \WI \ in the atmosphere of TOI-1518 b, with the last two species being detected for the first time ever in an exoplanetary atmosphere. Furthermore, splitting the transit into pre- and post-$T_0$ phases allowed us to probe the terminator asymmetries. We highlighted how some species like \MgI, \VII, \CrI, \SrII, and \BaI, produce a higher S/N in the evening limb of the planet, contrary to other species such as \CaI, and \WI \ which dominate in the morning limb. While these phase-resolved variations require further high-precision follow-up to be definitively decoupled from potential NLTE effects or 3D wind patterns, they strongly point toward a complex atmospheric stratification and dynamics, likely driven by equatorial jets, localised cold traps, and global circulation winds. Additionally, we investigated the possibility of constraining the stellar spin axis inclination ($i_\star$), and consequently the true 3D obliquity of the system, by searching for the stellar rotation period ($P_{\rm rot}$) in the out-of-transit TESS photometry. Given $v \sin i_\star \approx 79.7$ km s$^{-1}$ and $R_\star \approx 1.95$ $R_\odot$, the maximum expected rotation period is $\sim 1.25$ days. We computed a generalised Lomb-Scargle \citep[GLS,][]{Zechmeister2009} periodogram on the transit-masked light curve, performing a pre-whitening procedure to filter out the planetary proximity effects ($P_{\rm orb}$ and $P_{\rm orb}/2$). However, no statistically significant frequency emerged in the residuals. This non-detection is likely due to the lack of strong photospheric activity (e.g., starspots) capable of producing measurable photometric modulations, as expected for early-type, rapidly rotating stars. Consequently, a formal constraint on $i_\star$ via photometry currently remains out of reach. To place TOI-1518 b within the broader context of the known exoplanetary demographic, we compared its physical and orbital parameters with the confirmed planets from the NASA Exoplanet Archive (Fig.~\ref{fig:distribution}). In terms of its bulk physical properties, TOI-1518 b occupies the extreme parameter space characteristic of the UHJ population. Its high mass ($\approx 2.1$ $M_{\rm Jup}$) and low bulk density, coupled with the extreme insolation flux received from its early-type (hot) host star, firmly place it among the most inflated gas giants known to date. The system resides in the upper-right quadrant of the stellar effective temperature versus planetary equilibrium temperature plane, a regime where the atmospheric chemistry is expected to be dominated by atomic and ionic species due to the thermal dissociation of molecules. Specifically, with an updated $T_{\rm eq} \approx 2580$~K and a massive atmospheric scale height of $H \approx 609$~km (corresponding to $\sim 0.45\%$ of the planetary radius), TOI-1518 b represents one of the most accessible targets for transmission spectroscopy.
While its physical properties are consistent with the general population of short-period, highly irradiated giants, its dynamical architecture sets it significantly apart. As highlighted in the obliquity versus inclination plane (bottom-right panel of Fig.~\ref{fig:distribution}), TOI-1518 b resides in an isolated region of the parameter space, far from the bulk of well-aligned planetary systems. Its strongly misaligned, nearly polar orbit ($\lambda \approx 238\,deg$) represents a distinct dynamical signature. This extreme orbital configuration strongly points towards a history of violent dynamical interactions, such as high-eccentricity tidal migration driven by planet-planet scattering or Kozai-Lidov cycles, rather than a peaceful, disk-driven inward migration \citep{2026A&A...709L..15M}. Although a preliminary analysis of the \textit{Gaia} Renormalised Unit Weight Error (RUWE) does not currently suggest the presence of massive unresolved companions, their existence cannot be entirely ruled out. Further exploration of the possible companions existence are beyond the scope of the paper. 

\section{Conclusions}
\label{sec:conclusions}
In this work, we presented a comprehensive analysis of the UHJ TOI-1518 b (BD+66 1610\,b). By exploiting a joint Bayesian framework, we simultaneously modelled transit photometry and high-resolution spectroscopy (HARPS-N and SOPHIE) to bridge the gap between the system's dynamical architecture and its atmospheric composition.

Our joint dynamical and spectroscopic modelling confirmed the strongly misaligned, nearly polar architecture of the system and provided a robust empirical detection of orbital nodal precession, validating previous findings for the target. By anchoring the stellar rotation profile and mitigating the photospheric contamination, we successfully broke the system's geometrical degeneracies, retrieving a global atmospheric blue-shift, indicative of strong day-to-night side winds at the planetary terminator.

The wind velocity fitted globally during the orbital and RM modelling matches the mean kinematic shift extracted independently through our cross-correlation analysis. We reported the detection of several atomic and ionic species:  \HI,  \CI,  \NaI,  \CaI,  \ScII,  \TiII,  \VII,  \CrI,  \FeI,  \FeII,  \NiI,  \SrII,  \YII,  \EuII,  and \WI. Among these, \HI, \CI, \ScII, \VII, \SrII, \YII, \EuII, and \WI\ are detected for the first time in this planet, and \EuII\ and \WI\ for the first time in an exoplanetary atmosphere to our knowledge. Finally, our phase-resolved approach revealed distinct chemical asymmetries between the morning and evening limbs. These variations strongly point toward a highly stratified, three-dimensional atmospheric circulation, likely shaped by localised condensation phenomena and equatorial jets.

In conclusion, TOI-1518 b stands out as a prime laboratory to study the interplay between extreme dynamical histories and complex atmospheric physics. The synergy between orbital tomography and transmission spectroscopy presented in this work proved crucial to disentangle genuine atmospheric signals from stellar contamination in the fast-rotating regime. Future high-precision, broad-band observations (e.g., with JWST or the upcoming \textit{Ariel} mission) will be vital to probe the deeper atmospheric layers and fully constrain the morning-to-evening terminator chemical gradients that our optical CCF analysis has begun to unveil.

\section*{Data availability}
The HARPS-N radial velocities are only available in electronic form at the
CDS via anonymous ftp to \url{cdsarc.u-strasbg.fr} (130.79.128.5) or via
\url{http://cdsweb.u-strasbg.fr/cgi-bin/qcat?J/A+A/}.

\begin{acknowledgements}
Based on observations made with the Italian Telescopio Nazionale {\it Galileo} (TNG) operated by the {Fundaci\'on Galileo Galilei} (FGG) of the {Istituto Nazionale di Astrofisica} (INAF) at the {Observatorio del Roque de los Muchachos} (La Palma, Canary Islands, Spain). The authors sincerely thank the Referee for their comments that we think improved the quality of the manuscript. The authors acknowledge financial contribution from the INAF GO Large Grant 2023 GAPS-2, from the INAF GO Large Grant 2024 BRIDGES as well as from the European Union - Next Generation EU RRF M4C2 1.1 PRIN MUR 2022 project 2022CERJ49 (ESPLORA). Part of the research activities described in this paper were carried out with contribution of the Next Generation EU funds within the National Recovery and Resilience Plan (PNRR), Mission 4 - Education and Research, Component 2 - From Research to Business (M4C2), Investment Line 3.1 - Strengthening and creation of Research Infrastructures, Project IR0000034 – “STILES - Strengthening the Italian Leadership in ELT and SKA”. The authors acknowledge the Italian center for Astronomical Archives (IA2, https://www.ia2.inaf.it), part of the Italian National Institute for Astrophysics (INAF), for providing technical assistance, services, and support for GAPS collaboration activities. E.H.-C. acknowledges support from grant PRE2020-094770 under project PID2019-109522GB-C51, and project PID2022-137241NB-C42, funded by the Spanish Ministry of Science, Innovation and Universities / State Agency of Research, and by ‘ERDF, A way of making Europe’. T.Zi acknowledges support from CHEOPS ASI-INAF agreement n. 2019-29-HH.0, the NVIDIA Academic Hardware Grant Program for the use of the Titan V GPU card, and the Italian MUR Departments of Excellence grant 2023–2027 ‘Quantum Frontiers’. This work has made use of the VALD database, operated at Uppsala University, the Institute of Astronomy RAS in Moscow, and the University of Vienna. This paper includes data collected with the TESS mission, obtained from the MAST data archive at the Space Telescope Science Institute (STScI). Funding for the TESS mission is provided by the NASA Explorer Program. STScI is operated by the Association of Universities for Research in Astronomy, Inc., under NASA contract NAS 5–26555.\\
\end{acknowledgements}

\bibliographystyle{aa} % style aa.bst
\bibliography{bibliography}

\begin{appendix}
\section{Constraint on atmospheric trace modelling}
\label{app:sigma}
\begin{figure}
    \centering
    \includegraphics[width=\linewidth]{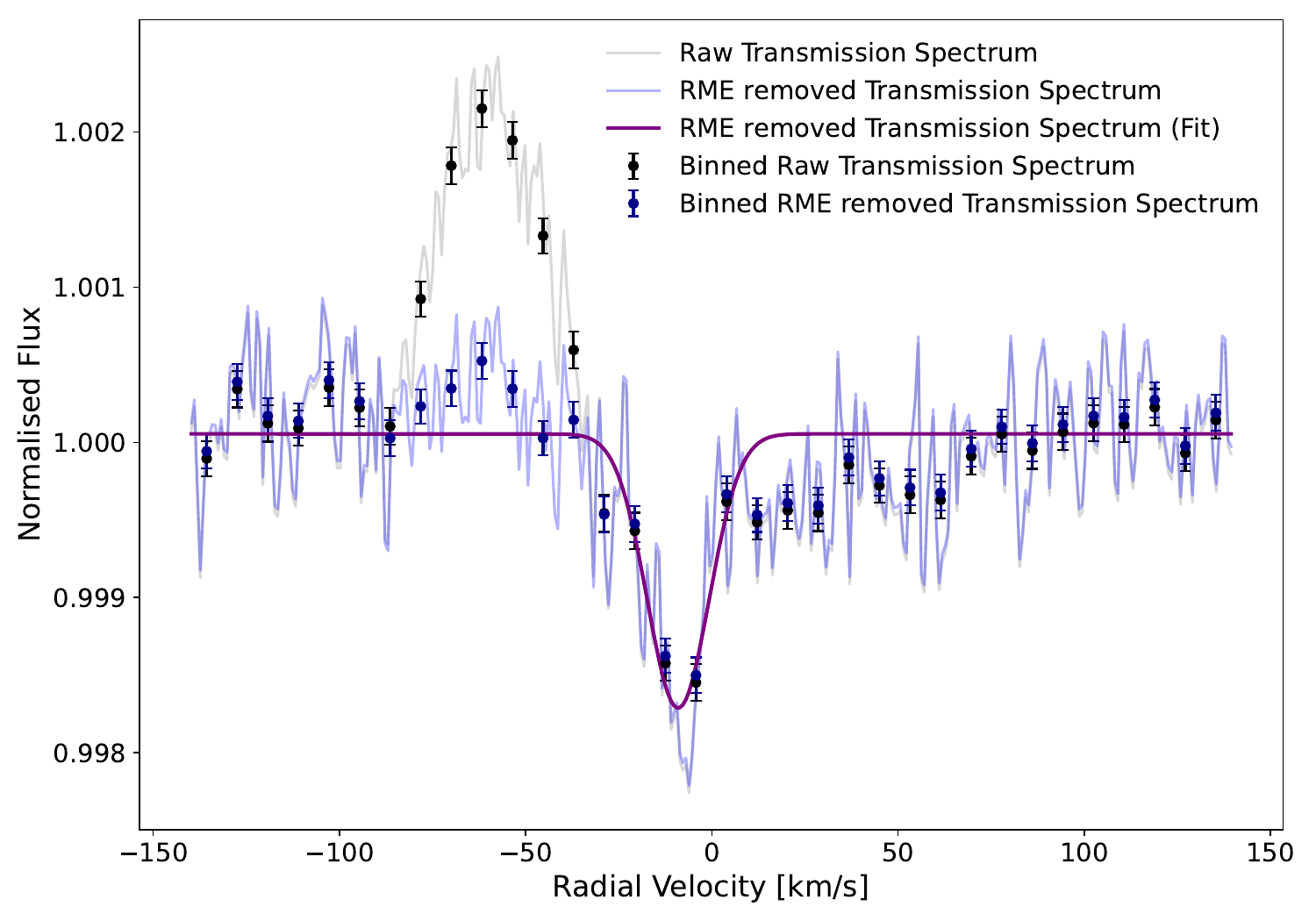}
    \caption{One-dimensional transmission spectra extracted from the tomographic LSD residuals in the planetary rest frame. The gray curve shows the spectrum prior to the removal of the RM effect, highlighting the apparent emission effect due to the Doppler Shadow. The blue curve displays the cleaned transmission spectrum after the RM effect subtraction. The purple solid line represents the Gaussian fit used to accurately measure the intrinsic width of the planetary track ($\sigma_2$), which provides the empirical prior for the joint MCMC analysis.}
    \label{fig:sigma_limit}
\end{figure}

During the joint Bayesian modelling, the determination of the width of the planetary atmospheric track ($\sigma_2$) in the tomographic maps can be affected by systematic residuals. In fast-rotating stars like TOI-1518, the regions surrounding the Doppler shadow often exhibit slight absorption or emission excesses. These features arise from flux conservation constraints during the normalisation of the spectral profiles and from small discrepancies in the RM effect modelling. If left unconstrained, the MCMC sampler may over-estimate the planetary $\sigma_2$ by incorporating these nearby stellar residuals into the planetary Gaussian profile. To mitigate this effect and ensure a physically sound retrieval, we performed a preliminary analysis. We first modelled and removed the RM effect from the 2D tomographic data using the best-fit parameters from a standard iteration. We then extracted the 1D transmission spectrum in the planetary rest frame and fitted the resulting absorption feature with a stand-alone Gaussian profile. The width retrieved from this "cleaned" 1D spectrum was then used to set a narrow Gaussian prior on the $\sigma_2$ parameter during the final joint global fit. In Fig.~\ref{fig:sigma_limit}, we show the gaussian fit of the RM effect removed transmission spectrum along with the raw one.

\section{Additional Figures and Tables}

\begin{figure}
    \centering
    \includegraphics[width=\linewidth]{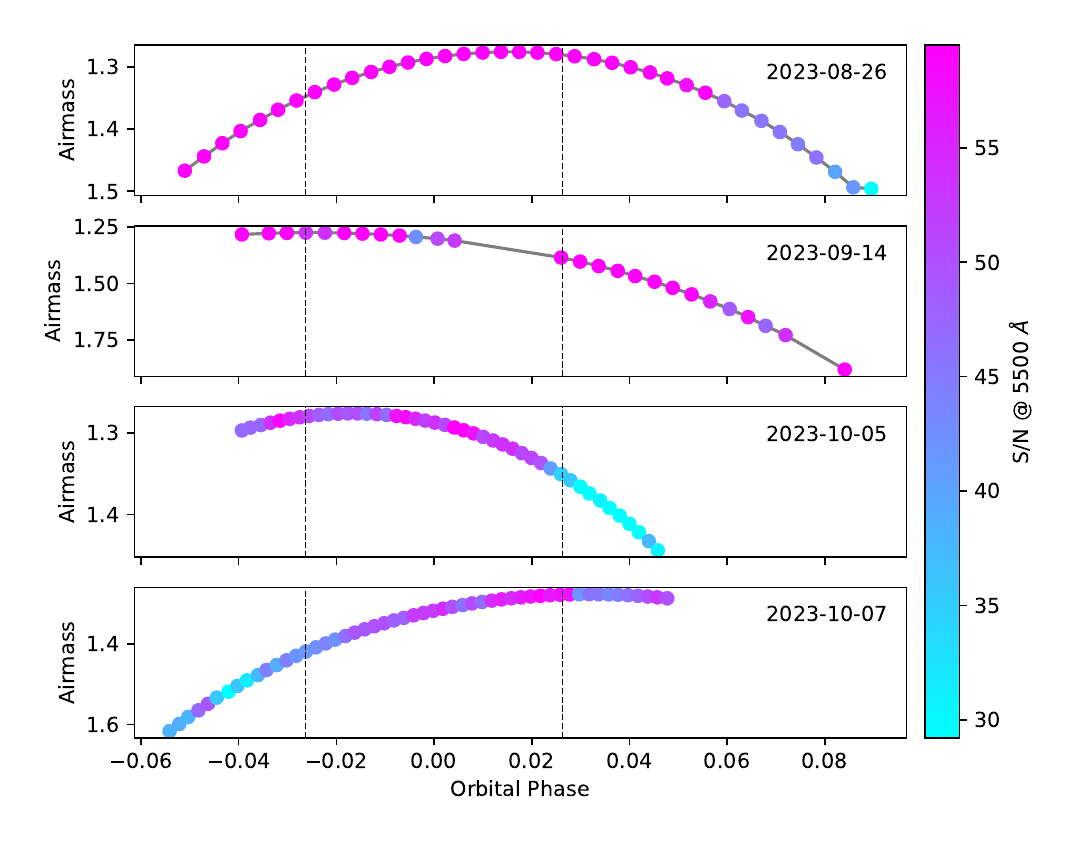}
    \caption{Airmass as a function of the orbital phase for the observed nights of TOI-1518 b. Dashed vertical lines represent the ingress and egress, while colourmap shows the S/N.}
    \label{fig:toi1518_obs}
\end{figure}

\begin{table*}
    \centering
    \caption{System parameters of TOI-1518.}
    \renewcommand{\arraystretch}{1.2} 
    \begin{tabular}{lccc}
        \hline\hline
        Parameter & Prior & Value & Reference\\
        \hline
        \multicolumn{4}{c}{Adopted and fixed parameters} \\
        \hline
        Stellar mass, $M_\star$ ($M_\sun$) & ... & 1.79 $\pm$ 0.26 & \cite{cabot21} \\
        Stellar radius, $R_\star$ ($R_\sun$) & ... & 1.95 $\pm$ 0.05 &  \cite{cabot21} \\
        Effective temperature, $T_{\rm eff}$ (K) & ... & 7300 $\pm$ 100 & \cite{cabot21} \\
        Surface gravity, $\log{g}$ (cgs) & ... & 4.1 $\pm$ 0.2 & \cite{cabot21} \\  
        Metallicity, [Fe/H] (dex) & ... & -0.10 $\pm$ 0.12 & \cite{cabot21} \\ 
        Eccentricity, $e$ & ... & 0 & Fixed \\
    
        \hline
        \multicolumn{4}{c}{Fitted parameters} \\
        \hline
        Orbital Period, $P$ (days) & $\mathcal{G}(1.90261, 1.90261)$ & 1.90261227$^{+0.00000018}_{-0.00000018}$ & This work \\
        Time of transit, $T_c$ (BJD) & $\mathcal{G}(2.46018e+06, 2.46018e+06)$ & 2460183.566347$^{+0.000028}_{-0.000029}$ & This work \\
        TESS Linear LDC, $\epsilon_t$ & $\mathcal{G}(0.206, 0.606)$ & 0.276$^{+0.022}_{-0.022}$ & This work \\
        HARPS-N Linear LDC, $\epsilon_h$ & $\mathcal{G}(0.377, 0.777)$ & 0.654$^{+0.025}_{-0.030}$ & This work \\
        Core radius ratio, $R_p/R_\star$ & $\mathcal{U}(0.05, 0.1598)$ & 0.09892$^{+0.00033}_{-0.00033}$ & This work \\
        Projected rotational velocity, $v \sin i_\star$ (km s$^{-1}$) & $\mathcal{U}(70.9, 86.9)$ & 79.70$^{+0.19}_{-0.47}$ & This work \\
        Atmospheric RM effect amplitude, $(R_p/R_\star)_{\rm atmo}$ & $\mathcal{U}(0.02, 0.1)$ & 0.0794$^{+0.0084}_{-0.0091}$ & This work \\
        Reference inclination at $T_c$, $i_0$ (deg) & $\mathcal{U}(77.24, 78.44)$ & 77.3029$^{+0.0085}_{-0.0083}$ & This work \\
        Nodal precession rate, $\mathrm{d}i/\mathrm{d}t$ (deg yr$^{-1}$) & $\mathcal{U}(-3.6525, 3.6525)$ & 0.1530$^{+0.0036}_{-0.0035}$ & This work \\
        Projected obliquity, $\lambda$ (deg) & $\mathcal{U}(220.34, 260.34)$ & 238.21$^{+0.43}_{-0.30}$ & This work \\
        Stellar RV semi-amplitude \tablefootmark{a}, $K$ (m s$^{-1}$) & $\mathcal{U}(0.1, 392)$ & 206$^{+42}_{-43}$ & This work \\
        Stellar RV semi-amplitude \tablefootmark{b}, $K$ (m s$^{-1}$) & $\mathcal{U}(0.1, 392)$ & 220$^{+29}_{-29}$ & This work \\
        Planetary RV semi-amplitude, $K_p$ (km s$^{-1}$) & $\mathcal{U}(105, 305)$ & 188.1$^{+6.6}_{-6.6}$ & This work \\
        Atmospheric wind, $v_{\rm w}$ (km s$^{-1}$) & $\mathcal{U}(-10, 2)$ & -7.02$^{+0.42}_{-0.42}$ & This work \\
        \hline
        \multicolumn{4}{c}{Derived parameters} \\
        \hline
        Systemic velocity, $V_{\mathrm{sys}}$ (\kms) & ... & -10.103 $\pm$ 0.074 & This work\\ 
        Semi-major axis, $a$ (AU) & ... & 0.0365$^{+0.0017}_{-0.0018}$ & This work \\
        Scaled semi-major axis, $a/R_\star$ & ... & 4.02$^{+0.21}_{-0.23}$ & This work \\
        Impact parameter, $b$ & ... & 0.884$^{+0.047}_{-0.049}$ & This work \\
        Impact parameter rate, $\mathrm{d}b/\mathrm{d}t$ (yr$^{-1}$) & ... & -0.01047$^{+0.00063}_{-0.00062}$ & This work \\
        Planetary mass, $M_p$ ($M_{\rm Jup}$) & ... & 2.17$^{+0.46}_{-0.41}$ & This work \\
        Planetary radius, $R_p$ ($R_{\rm Jup}$) & ... & 1.877$^{+0.048}_{-0.048}$ & This work \\
        Planetary density, $\rho_p$ (g cm$^{-3}$) & ... & 0.408$^{+0.093}_{-0.081}$ & This work \\
        Surface gravity, $g_p$ (m s$^{-2}$) & ... & 15.3$^{+3.3}_{-3.0}$ & This work \\
        Equilibrium temperature, $T_{\rm eq}$ (K) & ... & 2574$^{+83}_{-75}$ & This work \\
        Atmospheric scale height, $H$ (km) & ... & 609$^{+151}_{-112}$ & This work \\
        Fractional scale height, $H/R_p$ (\%) & ... & 0.45$^{+0.11}_{-0.08}$ & This work \\

        \hline
    \end{tabular}
    \tablefoot{  The values are divided into three blocks: parameters adopted from literature or fixed in our analysis, parameters dynamically fitted in our joint MCMC framework, and derived physical quantities. For the fitted parameters, the adopted priors are listed as $\mathcal{U}(a, b)$ for uniform distributions and $\mathcal{G}(a, b)$ for Gaussian distributions bounded between $a$ and $b$, with mean $\mu$ and standard deviation $\sigma$.
        \tablefoottext{a}{Only HARPS-N.}
        \tablefoottext{b}{HARPS-N + SOPHIE. For the derived parameters we employed the Stellar RV semi-amplitude retrieved from the HARPS-N + SOPHIE joint fit.}
    }

    \label{tab:toi1518_par}
\end{table*}

\begin{table*}[ht]
\caption{Log of the transit observations of TOI-1518 b used in this work.} 
\centering
\begin{tabular}{cccccccc}
\hline\hline 
Night & Date & Programme & P.I. & \# Spectra & S/N @ 5500\,\AA & Exposure & Airmass \\ 
 & & & & (In/Out of transit) & & time [s] & \\ \hline
1 & 26 August 2023 & CAT23A\_69 & Herrero Cisneros & 38 (14/24) & 62 & 600 & 1.276--1.496 \\
2 & 14 September 2023 & CAT23A\_69 & Herrero Cisneros & 26 (10/16) & 62 & 600 & 1.276--1.880 \\
3 & 5 October 2023 & BRIDGES & Borsa & 44 (27/17) & 46 & 300 & 1.276--1.444 \\
4 & 7 October 2023 & BRIDGES & Borsa & 52 (27/25) & 47 & 300 & 1.276--1.617 \\
\hline
\end{tabular}
\label{tab:observations}
\end{table*}

\begin{figure}
    \centering
    \includegraphics[width=0.8\linewidth]{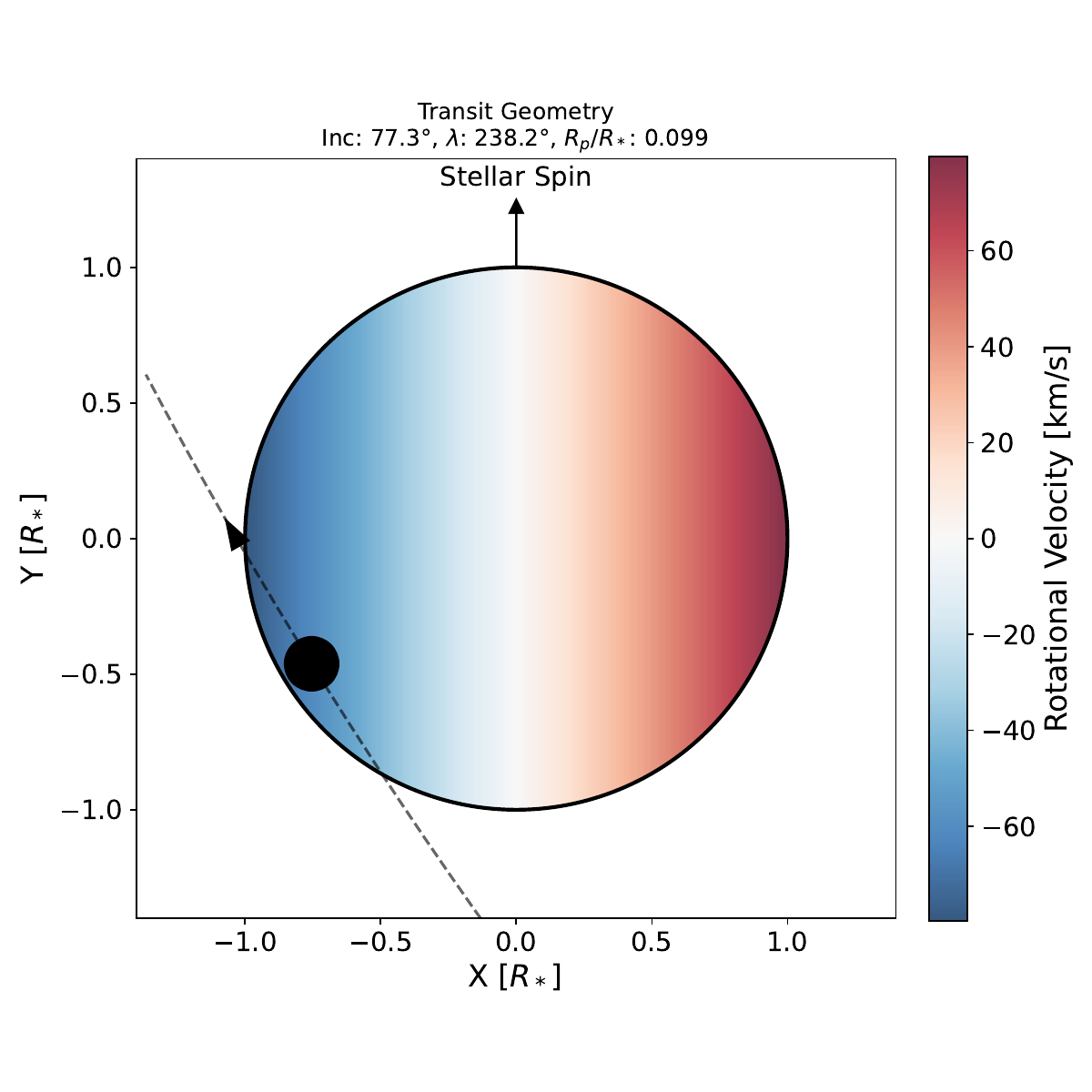}
    \caption{Transit geometry obtained with the parameters retrieved from our analysis.}
    \label{fig:transit_geometry}
\end{figure}

\begin{figure}
    \centering
    \includegraphics[width=\linewidth]{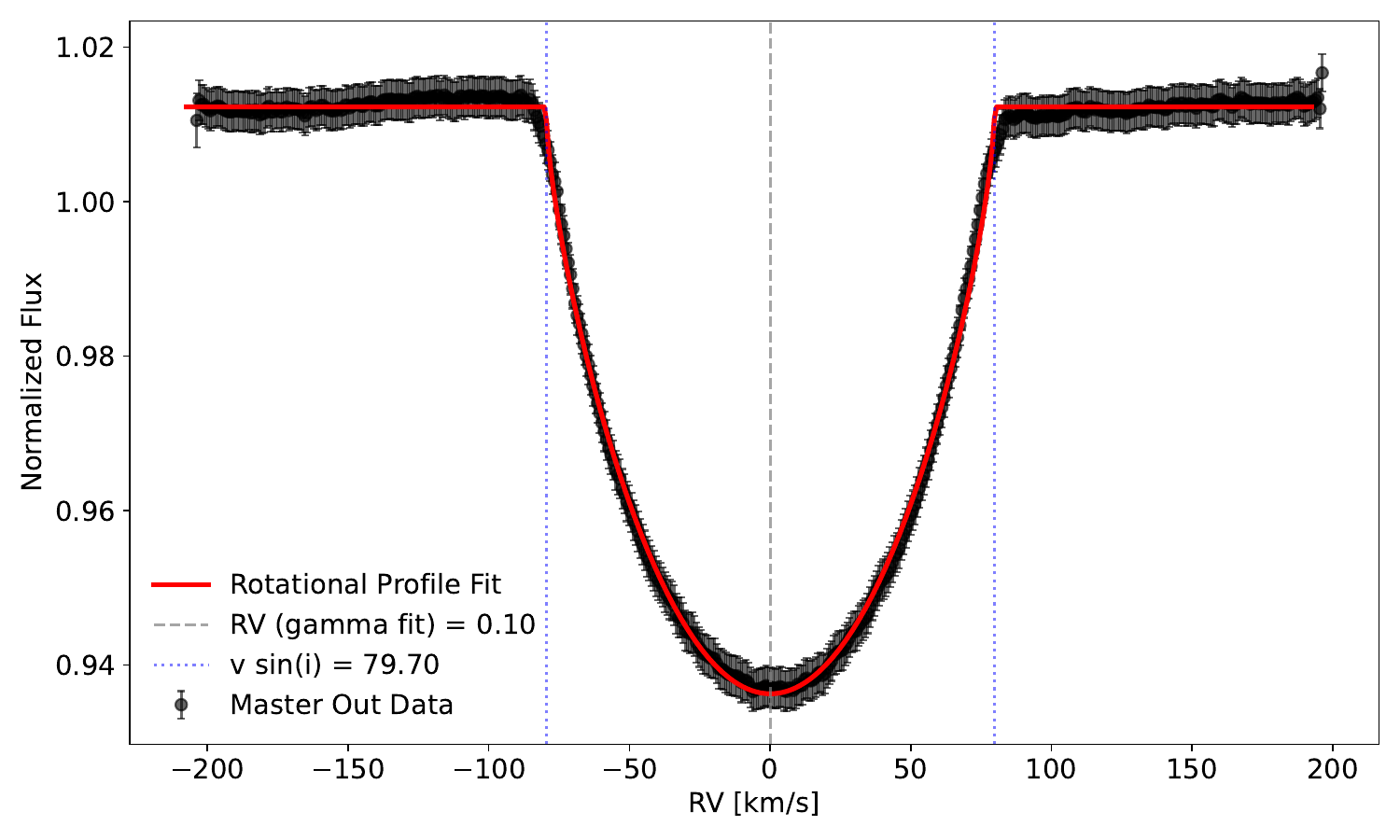}
    \caption{Average master-out LSD profile with \cite{Gray2005} model.}
    \label{fig:prof}
\end{figure}

\begin{table*}[t]
    \centering
    \caption{Results of the single-line transmission spectroscopy analysis.}
    \begin{tabular}{lcccc}
        \hline\hline
        Line & Absorption & FWHM & $v_{\rm wind}$ & $R_{\rm eff}/R_{\rm p}$ \\
             & (\%)       & (km s$^{-1}$) & (km s$^{-1}$) & \\
        \hline
        H$\alpha$ & 1.93 $\pm$ 0.17 & 19.7 $\pm$ 2.1 & --4.84 $\pm$ 0.84 & 1.75 $\pm$ 0.05 \\
        Na\,\textsc{i} D1 & 0.34 $\pm$ 0.10 & 19.7 $\pm$ 6.7 & +8.1 $\pm$ 2.8 & 1.17 $\pm$ 0.05 \\
        Na\,\textsc{i} D2 & 0.53 $\pm$ 0.10 & 17.3 $\pm$ 4.1 & --7.2 $\pm$ 1.7 & 1.25 $\pm$ 0.04 \\
        Fe\,\textsc{ii} $\lambda$5018.43\,\AA & 0.70 $\pm$ 0.11 & 19.9 $\pm$ 3.7 & --8.4 $\pm$ 1.5 & 1.32 $\pm$ 0.04 \\
        Fe\,\textsc{ii} $\lambda$5169.02\,\AA & 0.78 $\pm$ 0.11 & 28.5 $\pm$ 4.9 & --9.2 $\pm$ 1.9 & 1.36 $\pm$ 0.04 \\
        \hline
    \end{tabular}
    \label{tab:single_line_results}
\end{table*}

\begin{figure}
    \centering
    \includegraphics[width=\linewidth]{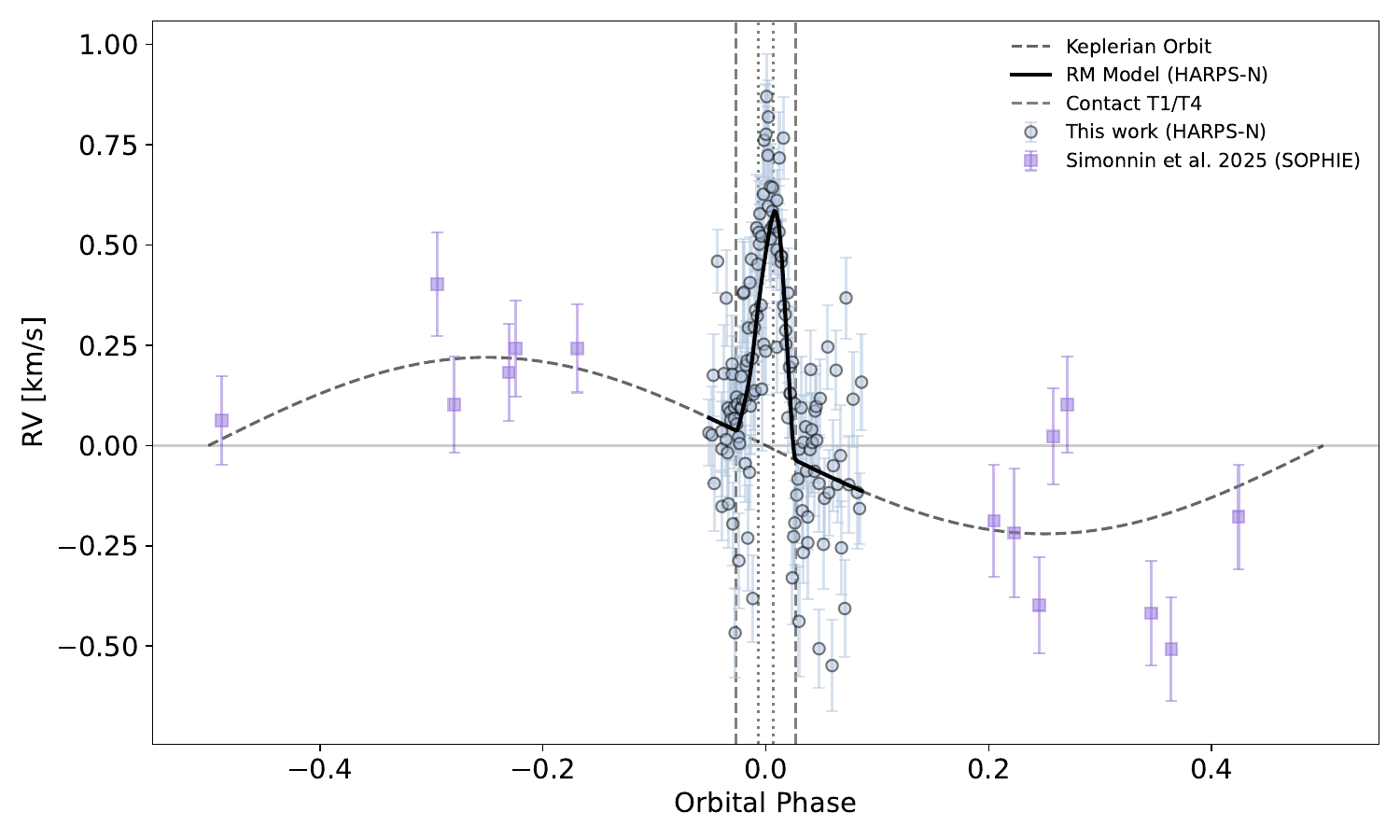}
    \caption{SOPHIE + HARPS-N RVs with fitted model.}
    \label{fig:full_rv}
\end{figure}

\begin{figure*}
    \centering
    \includegraphics[width=\linewidth]{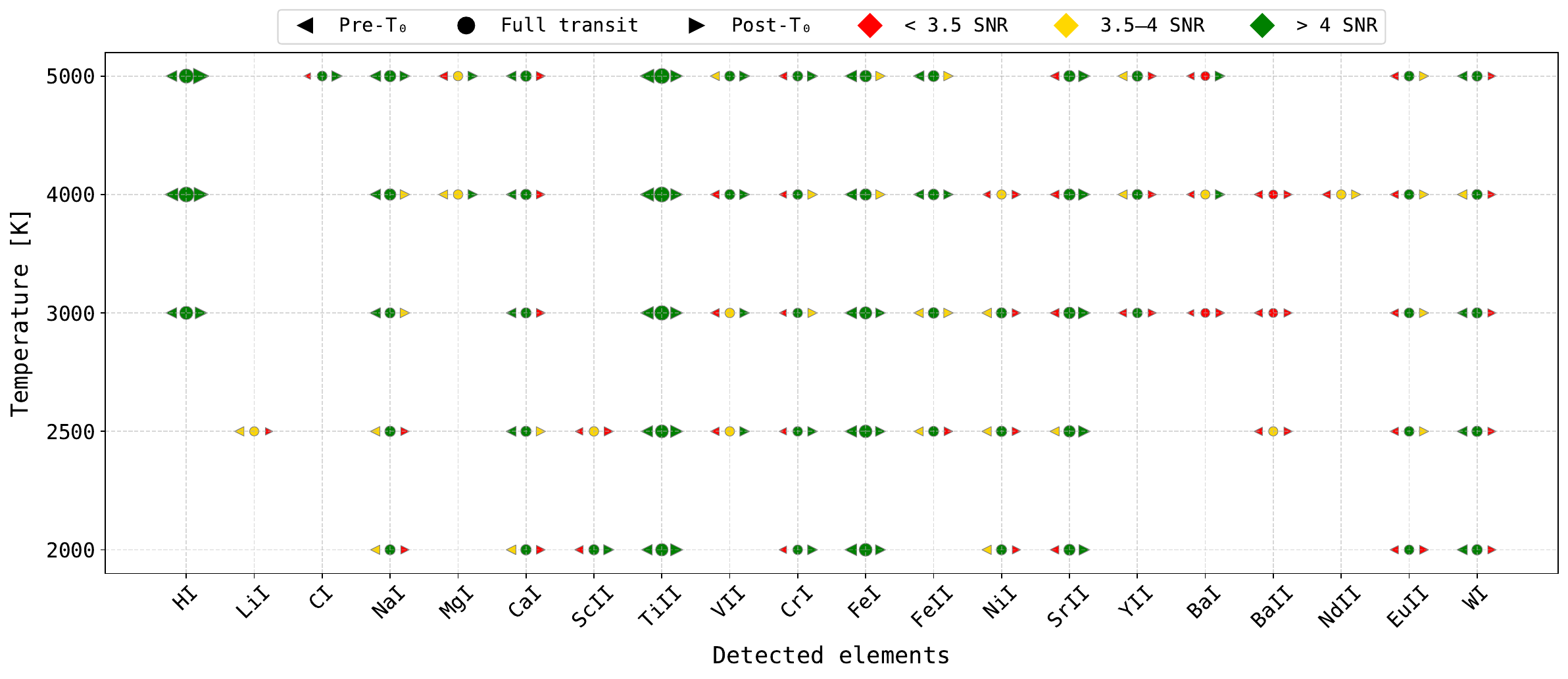}
    \caption{Summary of the CCF analysis results. Marker sizes scale with the amplitude derived from the Gaussian+offset fit to the one-dimensional $K_p$–$V_{\mathrm{sys}}$ profiles, computed for the full transit as well as for the pre- and post-$T_0$ phases. The colour coding reflects the detection significance: green indicates robust detections (S/N $>$ 4), yellow corresponds to marginal detections (3.5 $<$ S/N $<$ 4), and red denotes non-detections (S/N $\leq$ 3.5). Missing markers indicate species for which the $K_p$–$V_{\mathrm{sys}}$ signal did not meet the initial $3.5 \sigma$ threshold, and then they were excluded from the Gaussian fitting procedure.}
    \label{fig:kpvsys_fit}
\end{figure*}

\begin{table*}[t]
    \centering
    \caption{Maximum full-transit S/N values for each detected element, together with the corresponding pre- and post-$T_0$ S/Ns.}
    \renewcommand{\arraystretch}{1.1}
    \fontsize{10}{15}\selectfont
    \begin{tabular}{lccccc}
    \hline
    \hline
     & Full transit S/N & RV [\kms] & $K_p$ [\kms] & Pre-$T_0$ S/N & Post-$T_0$ S/N \\
    \hline
    \HI   & $11.18 \pm 0.22$ & $-6.24 \pm 0.21$ & $182 \pm 23$       & $7.87 \pm 0.39$ & $9.99 \pm 0.27$ \\  
    \LiI  & $3.53 \pm 0.53$  & $-6.69 \pm 0.90$ & $193 \pm 20$       & $3.72 \pm 0.35$ & $1.92 \pm 0.85$ \\
    \CI   & $4.31 \pm 0.65$  & $-0.28 \pm 0.49$ & $246 \pm 16$       & $1.32 \pm 0.23$ & $5.04 \pm 0.63$ \\
    \NaI  & $6.37 \pm 0.38$  & $-6.95 \pm 0.65$ & $240 \pm 21$       & $5.39 \pm 0.46$ & $3.93 \pm 0.19$ \\
    \MgI  & $3.85 \pm 0.27$  & $-10.65 \pm 0.65$& $181 \pm 79$       & $3.53 \pm 0.33$ & $4.02 \pm 0.30$ \\
    \CaI  & $5.37 \pm 0.39$  & $-7.28 \pm 0.35$ & $213 \pm 19$       & $4.50 \pm 0.44$ & $3.47 \pm 0.43$ \\   
    \ScII & $4.76 \pm 0.73$  & $-5.40 \pm 0.35$ & $223 \pm 29$       & $2.86 \pm 0.41$ & $4.58 \pm 0.81$ \\
    \TiII & $11.41 \pm 0.39$ & $-8.10 \pm 0.18$ & $196 \pm 11$       & $9.40 \pm 0.47$ & $6.68 \pm 0.39$ \\
    \VII  & $4.58 \pm 0.58$  & $-9.08 \pm 0.41$ & $167 \pm 12$       & $3.52 \pm 0.52$ & $4.57 \pm 0.54$ \\    
    \CrI  & $4.45 \pm 0.35$  & $-9.77 \pm 0.51$ & $241 \pm 13$       & $2.56 \pm 0.38$ & $4.76 \pm 0.45$ \\
    \FeI  & $7.88 \pm 0.40$  & $-8.60 \pm 0.33$ & $199.6 \pm 7.8$    & $6.30 \pm 0.48$ & $4.45 \pm 0.31$ \\   
    \FeII & $5.67 \pm 0.36$  & $-6.74 \pm 0.47$ & $181 \pm 27$       & $5.02 \pm 0.31$ & $3.93 \pm 0.43$ \\    
    \NiI  & $5.02 \pm 0.67$  & $-6.04 \pm 0.30$  & $175.1 \pm 7.9$    & $3.69 \pm 0.77$ & $2.86 \pm 0.44$ \\    
    \SrII & $6.35 \pm 0.26$  & $-7.39 \pm 0.37$ & $225 \pm 15$       & $3.41 \pm 0.34$ & $6.67 \pm 0.31$ \\
    \YII  & $4.85 \pm 0.58$  & $-3.38 \pm 0.41$ & $261 \pm 13$       & $3.72 \pm 0.69$ & $2.7 \pm 1.1$ \\
    \BaI  & $3.58 \pm 0.66$  & $-11.18 \pm 0.42$& $129 \pm 16$       & $2.18 \pm 0.80$ & $4.24 \pm 0.72$ \\
    \BaII & $3.65 \pm 0.44$  & $-11.23 \pm 0.71$& $212 \pm 19$       & $2.67 \pm 0.32$ & $2.94 \pm 0.60$ \\
    \NdII & $3.83 \pm 0.62$  & $-11.69 \pm 0.32$& $197 \pm 54$       & $2.86 \pm 0.90$ & $3.52 \pm 0.48$ \\
    \EuII & $4.39 \pm 0.57$  & $-7.25 \pm 0.54$ & $113 \pm 11$       & $2.76 \pm 0.72$ & $3.65 \pm 0.54$ \\
    \WI   & $5.01 \pm 0.71$  & $-1.90 \pm 0.30$ & $210.1 \pm 7.3$    & $4.40 \pm 0.70$ & $2.77 \pm 0.60$ \\  
    All   & $11.65 \pm 0.32$ & $-7.89 \pm 0.18$ & $194 \pm 11$       & $10.45 \pm 0.44$& $6.88 \pm 0.36$ \\
    \hline
    \end{tabular}
    \label{tab:detection}
\end{table*}

\begin{figure*}
    \centering
    \includegraphics[width=\linewidth]{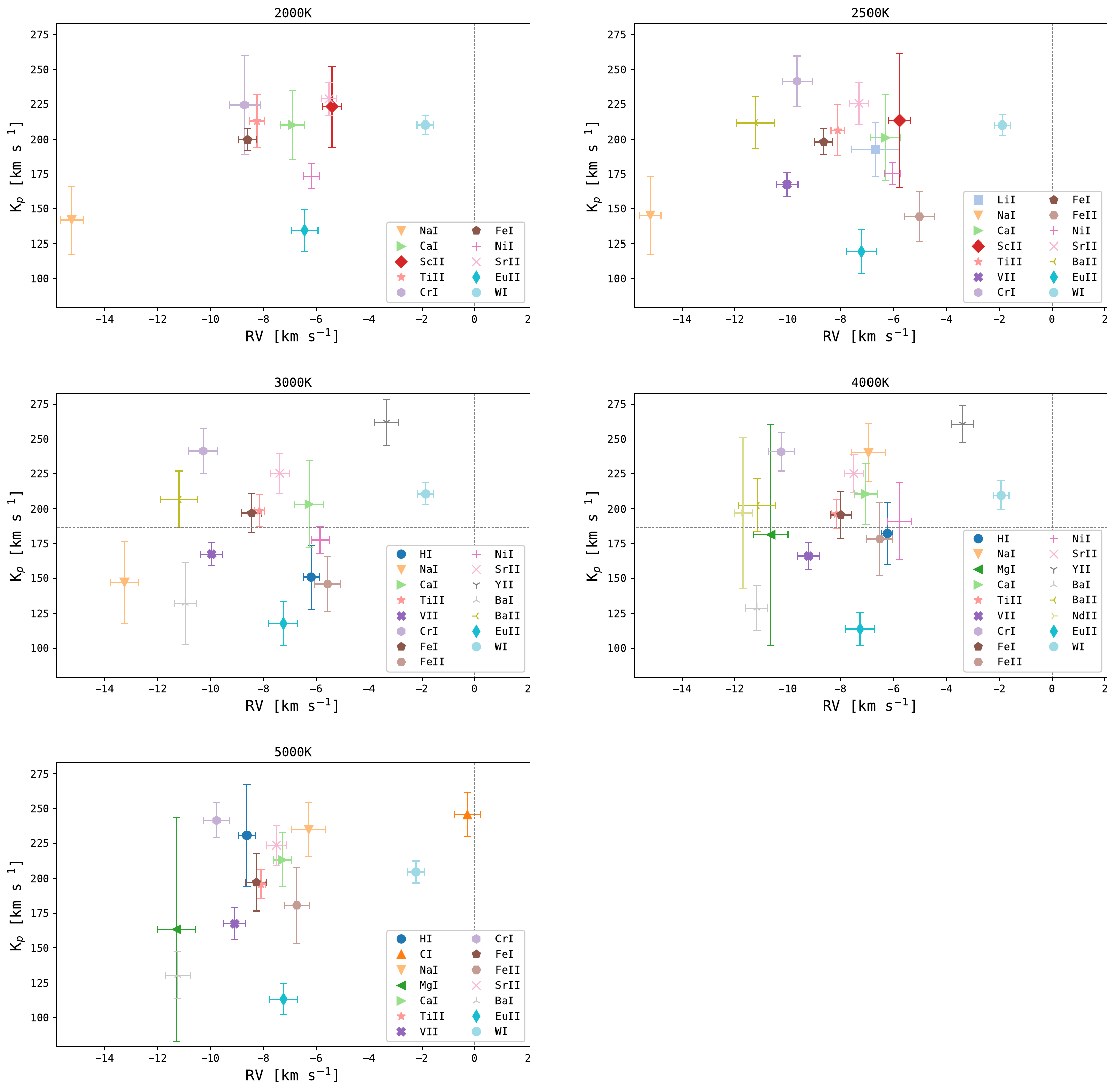}
    \caption{Same as Fig.~\ref{fig:KpRV}, but resolved individually for each of the five isothermal temperature templates (2000~K, 2500~K, 3000~K, 4000~K, and 5000~K) employed in the cross-correlation analysis. Each panel displays the retrieved RV centroids and $K_{\text{p}}$ values for the species that passed the detection or tentative threshold at that specific template temperature. The horizontal and vertical dashed lines, as well as the yellow bands representing the weighted mean values, are defined as in Fig.~\ref{fig:KpRV}.}
    \label{}
\end{figure*}

\begin{figure*}
    \centering
    \includegraphics[width=\linewidth]{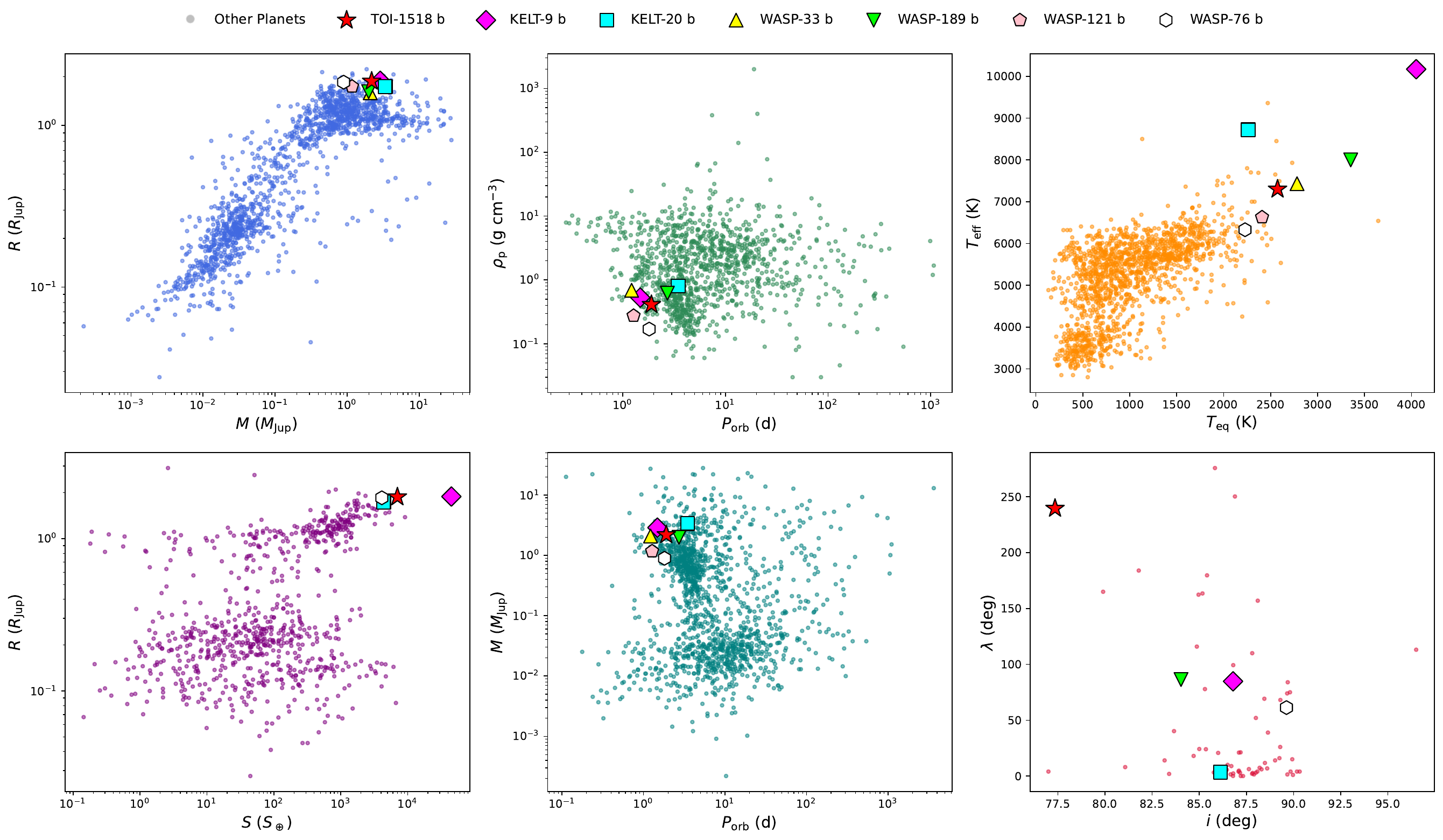}
    \caption{Planetary and stellar parameter distributions for TOI-1518 b within the context of the known exoplanet population. In all panels, gray dots represent the population of confirmed transiting exoplanets retrieved from the NASA Exoplanet Archive. TOI-1518 b is highlighted with a red star (using the parameters derived in this work), while other coloured symbols denote a comparison sample of well-known UHJs, including KELT-9 b, WASP-189 b, and WASP-33 b.}
    \label{fig:distribution}
\end{figure*}

\begin{figure*}
    \centering
    \includegraphics[width=\linewidth]{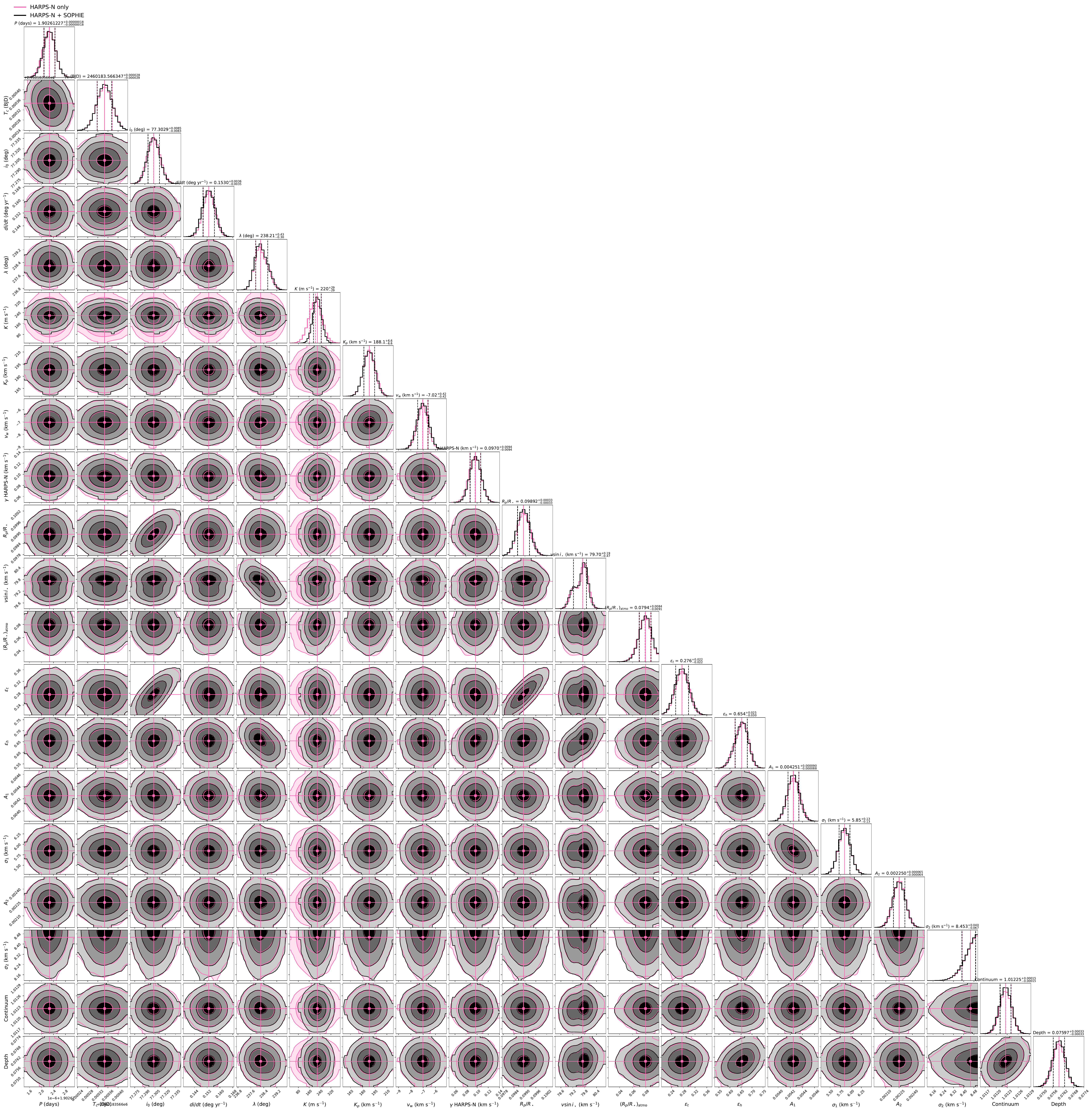}
    \caption{Corner plots of the analysis fitting the TESS, RV, MAP and PROF datasets without the literature RVs. In pink we show the distributions obtained employing only HARPS-N RVs, while in black we show the ones including SOPHIE data. The two distributions are almost identical except for the $K$, as expected.}
    \label{fig:corner}
\end{figure*}

\end{appendix}
\end{document}